\documentclass{aa}
\usepackage{graphicx}
\usepackage{txfonts}

\newcommand{\Mini}{\mbox{$M_{\rm i}$}}

\newcommand{\feh}{\mbox{\rm [{\rm Fe}/{\rm H}]}}

\newcommand{\oh}{\mbox{\rm [{\rm O}/{\rm H}]}}
\newcommand{\Msun}{\mbox{$M_{\odot}$}}
\newcommand{\Teff}{\mbox{$T_{\rm eff}$}}
\newcommand{\Te}{\mbox{$T_{\rm e}$}}
\newcommand{\Teo}{\mbox{$T_{\rm e}^0$}}
\newcommand{\ttr}{\mbox{$t_{\rm tr}$}}

\newcommand{\logL}{\mbox{$\log L/L_{\odot}$}}

\newcommand{\hii}{\mbox{H\,{\sc ii}}}
\newcommand{\comment}[1]{}
\newcommand{\forbline}[3]{\mbox{$[$#1\,{\sc #2}$]\,\lambda#3$}}
\newcommand{\recline}[3]{\mbox{#1\,{\sc #2}$\,\lambda#3$}}
\newcommand{\oiiiline}{\forbline{O}{iii}{5007}}
\newcommand{\heiiline}{\recline{He}{ii}{4686}}
\newcommand{\Hbeta}{\mbox{${\rm H}\beta$}}

\newcommand{\moiii}{\mbox{$m_{5007}$}}
\newcommand{\Moiii}{\mbox{$M_{5007}$}}
\newcommand{\muH}{\mbox{$\mu_{\rm H}$}}

\newcommand{\beq}{\begin{equation}}
\newcommand{\eeq}{\end{equation}}
\newcommand{\beqa}{\begin{eqnarray}}
\newcommand{\eeqa}{\end{eqnarray}}
\newcommand{\benu}{\begin{enumerate}}
\newcommand{\eenu}{\end{enumerate}}
\newcommand{\bite}{\begin{itemize}}
\newcommand{\eite}{\end{itemize}}
        \def\smallskip{\vskip 2pt}
\begin{document}

\title{Evolution of planetary nebulae}
\subtitle{II. Population effects on the bright cut-off of the PNLF}

\author{P. Marigo\inst{1} \and L. Girardi\inst{2} \and A. Weiss\inst{3} \and 
	M.A.T. Groenewegen\inst{4} \and C. Chiosi\inst{1}
}
\institute{
Dipartimento di Astronomia, Universit\`a di Padova,
	Vicolo dell'Osservatorio 2, I-35122 Padova, Italy \and
Osservatorio Astronomico di Trieste,
	Via Tiepolo 11, I-34131 Trieste, Italy \and
Max-Planck-Institut f\"ur Astrophysik, 
	Karl-Schwarzschild-Str. 1, Garching bei M\"unchen, Germany \and
PACS ICC-team, Instituut voor Sterrenkunde, Celestijnenlaan 200B,
	B-3001 Leuven, Belgium
	}

\offprints{Paola Marigo\\ \email{marigo@pd.astro.it} }

\date{To appear in Astronomy \& Astrophysics}

\abstract{
We investigate the  
bright cut-off of the \oiiiline\ planetary nebula luminosity function
(PNLF), that has been suggested as a powerful extragalactic distance 
indicator in virtue of its observed invariance 
against populations effects.
Theoretical PNLFs are constructed via Monte-Carlo 
simulations of populations of PNe, 
whose individual properties are described with the aid of recent  
PN synthetic models (Marigo et al. 2001), coupled to 
a detailed photoionisation code (CLOUDY).
The basic dependences of the cut-off magnitude $M^*$ are then discussed.
%, namely: age of the stellar progenitors, metallicity, transition time, 
%and nuclear-burning regime of the central stars.
We find that :
(i) In galaxies with recent or ongoing star formation, the modelled
PNLF present $M^*$ values between $-4$ and $-5$, depending on
model details. These are very close to the observationally-calibrated
value for the LMC.
(ii) In these galaxies, the PNLF cut-off is produced by PNe
with progenitor masses of about 2.5 \Msun, while less massive stars
give origin to fainter PNe. As a consequence $M^*$ is expected to depend
strongly on the age of the last burst of star formation, 
dimming by as much as 5 magnitudes as we go from
young to 10-Gyr old populations. 
(iii) Rather than on the initial metallicity of a stellar population,
$M^*$ depends on the actual [O/H] of the observed PNe,
a quantity that may differ significantly from the initial value
(due to dredge-up episodes),
especially in young and intermediate-age PN populations.
(iv) Also the transition time from the end of AGB to the PN phase,
and the nuclear-burning properties (i.e. H- or He-burning) of the 
central stars introduce non-negligible effects on $M^*$.
The strongest indication derived from the present calculations is a
 serious difficulty to explain the age-invariance of 
the cut-off brightness over an extended interval, say from 1 to 13 Gyr, 
that observations of PNLFs in galaxies of
late-to-early type seem to suggest.
We discuss the implications of our findings, also in relation
to other interpretative pictures proposed in the past literature.
%%%%%%%%%%%%%%%%%%%%%%%%%%%%%%%%
}

\titlerunning{Population effects in the PNLF}
\authorrunning{Marigo et al.}
\maketitle

%%%%%%%%%%%%%%%%%%%%%%%%%%%%%%%%%%%%%%%%%%%%%%%%%%%%%%%%%%%%%%%%%%%%%%%%%%
\section{Introduction}
\label{sec_intro}

The Planetary Nebula luminosity function (PNLF) constitutes one
of the most attractive indicators to set out the extragalactic
distance scale.
{From} the observational side, PNe can be identified in galaxies from 
the LMC to as far as the Virgo cluster, and are numerous in some 
of the key galaxies that link the ``local'' distance scale given by
Cepheids, with the more distant scale given by 
the surface brightness fluctuations (SBF) method, 
the Tully-Fisher relation, and SNe Ia. 
The PNLF is suitable for determining distances as far as 20--25~Mpc,
beyond which exposure times for reasonable S/N-ratios become prohibitively
long. 
The importance of the PNLF for the cosmological distance ladder is
fully described by e.g. Ciardullo (2003), Ciardullo \& Jacoby (1999) 
and Jacoby, Ciardullo \& Feldmeier (1999), Jacoby (1997), 
to whom we refer for all details. 

In practice, the PNLF as a candle works in the following way:
Once the flux in the \oiiiline\ emission line,
$F_{5007}$ (in cgs units),
is measured for a sample of PNe in an external galaxy, 
it is converted to \moiii\ magnitude,
\beq
\moiii = -2.5\,\log F_{5007} - 13.74
\label{eq_moiii}
\eeq
(Jacoby 1989). The number of objects at any given \moiii\ bin 
constitutes the PNLF, $N(\moiii)$, which can be fitted over the 
brightest magnitudes by the function
\beq
N(\moiii) = e^{0.307\,m_{5007}}
\left[ 1-e^{3(m^*-m_{5007})} \right]
\label{eq_cutoff}
\eeq
(see Ciardullo et al.~1989).
The key quantity is the cut-off magnitude $m^*$. According to
several arguments (see below), its corresponding absolute 
value, $M^*$, appears to be constant in most galaxies, hence 
providing an excellent (secondary) standard candle.
Therefore, a measure of $m^*$ gives a distance estimate in a 
single step, provided that interstellar reddening is suitably taken
into account. The present-day calibration
of $M^*$, based on a sample of galaxies with SBF and
Cepheid distances, and correcting for a small metallicity 
dependence, is $M^* = -4.47$ (Ciardullo et al. 2002).

The possibility of systematic variations in the value of
$M^*$ among galaxies has been investigated in several
different ways. From the observational side, the evidences for
a nearly constant maximum luminosity for PNe are numerous, as 
reviewed by e.g.\ Ciardullo (2003) and Jacoby (1997).
The main points are:
(1) $m^*$ does not significantly change with galactocentric radius in 
the galaxies M31, M81, NGC\,4494 and NGC\,5128, where significant
population gradients should be present. 
(2) There is no hint of large systematic variations in $m^*$ among 
galaxies of different types in the galaxy clusters/groups of 
M81, NGC\,1023, NGC\,5128,
Fornax, Leo~I, and Virgo. Overall, this is
indicating that age and metallicity variations 
have little effect on $M^*$. On the other hand, comparison with
Cepheid distances reveals a small non-monotonic dependence on
metallicity, expressed by the system's oxygen abundance, which
can be compensated for by a relation quadratic in [O/H] 
(Ciardullo et al.\ 2002).

Although these observational arguments seem compelling, 
they are not accompanied by a solid, unique theoretical framework. 
In other words, different authors 
do not agree on the reasons why $M^*$ should be constant,
reaching seemingly contradictory conclusions over the years. 
First, early theoretical arguments by Jacoby (1989) and 
PN modelling by Dopita et al.\ (1992) explain why the 
\oiiiline\ flux depends little on the PN metallicity.
At the same time, the PNLF simulations by M\'endez et al. (1993) 
indicate that a substantial dependence on population age might be
present, as indicated  by the differences in the best fits of the LMC and 
Milky Way PNLFs. Dopita et al. (1992) also suggest that
a dependence on age should be present. 
While M\'endez et al.\ (1993) claim that the bright end of the
PNLF is populated by optically-thin objects, Dopita et al.\ (1992)
model optically-thick objects only and, after metallicity and
population corrections, succeed in reproducing the PNLF and the
constant cutoff brightness. Similar open questions concern the
importance of hydrogen- (M\'endez et al.\ 1993) vs.\ helium-burning
(Ciardullo et al.\ 1989; Jacoby 1989) central stars.

More recently, the apparent lack of predicted population age
effects is explained by  Ciardullo \& Jacoby (1999) by
compensating circumstellar extinction. Young and 
massive central stars should be very luminous and result in 
over-luminous PNe (i.e. beyond the standard $M^*$), 
which is not observed. The proposed solution 
by Ciardullo \& Jacoby (1999) is  that, since the most massive  
asymptotic giant branch (AGB) stars likely 
become dust-enshrouded because of  heavy mass loss, 
the related circumstellar extinction 
would reduce the  \oiiiline\ emission from these objects, always 
to at least  $M^*$ or below.
Provided that PNe with lower-mass progenitors
are present, then the observed cut-off magnitude would be $M^*$. 
If this is the case, one has yet to understand which is the typical
age of the PNe that contribute to the cut-off, and explore the
metallicity effects in the relevant age intervals. 

The theoretical investigations of the PNLF, quoted above, have
different levels of sophistication in the models. Because of the
complexity of the problem, simplifications and severe
assumptions about the properties of PNe and their progenitor stars 
have always been introduced, like for instance:
an initial--final mass relation (hereinafter also IFMR) 
independent of metallicity
(e.g. M\'endez et al.\ 1993), either optically-thin or thick PNe
(Dopita et al.\ 1992; M\'endez et al.\ 1993; Stanghellini 1995;
M\'endez \& Soffner 1997), 
a narrow mass range being responsible for the bright end of the PNLF 
(Jacoby 1989; Dopita et al.\ 1992), and the restriction to either
H- or He-burner central stars only.
All these assumptions can be improved in some way, by using
the full results of stellar evolution and the population synthesis 
theory.

In the present paper, we elaborate on the theoretical 
framework for the PNLF predictions, based on our own simulations.
The starting
point for such a project has been the construction of an updated
synthetic code, that follows the PN evolution as a function of
basic stellar parameters, such as progenitor mass, metallicity, 
type of post-AGB evolution, etc. The complete code is 
described in Marigo et al. (2001; hereafter Paper~I), 
together with basic comparisons
with the available data for Galactic PNe. Note that this code is
much more complex than those previously used to investigate the
PNLF, as it takes into account  
the chemical composition and wind velocity of the ejected material 
as predicted by AGB models,
the interaction of the multiple winds as the PN expands,
the effect of shell thickening due to ionisation, etc. 
Moreover, there is no artificial assumption about 
the conditions of optical thickness of the nebulae.
The underlying model used to produce theoretical PNLFs is therefore
more realistic and much less dependent on explicit or hidden
assumptions than previous approaches. 
We will recapitulate the key features of our theoretical model in
Section~\ref{sec_overview}.

Despite all improvements, the Marigo et al.\ (2001) models
still treated some characteristics of the ionized nebulae in a
simplified way.
For instance, the electronic temperature \Te\ was introduced 
as a parameter and kept constant during the complete PN evolution.
The emission line fluxes were then derived by simple formulas 
based on this temperature,
on element abundances, and on the estimated sizes of the
Str\"omgren spheres for different ions. Since the flux of 
the optical forbidden lines -- and in particular the \oiiiline\ one
-- depends strongly  on \Te\ and the actual amount of ionized 
material, this simplification is clearly
inadequate for an investigation of the PNLF. Hence, in this paper
(Section~\ref{ssec_cloudy}) we relax the above-mentioned 
approximations by incorporating the photoionisation code CLOUDY 
(Ferland 2001) into our PN models, thereby improving them further,
and particularly towards the direction of accurate predictions of
spectral features.

Then, in Sect.~\ref{sec_samples} we describe the generation
of simulated PN samples in galaxies, firstly performing
some basic comparisons with the empirical data of Galactic PNe. 
We show how present models can reproduce,
to a large extent, some well-known properties of observed PNe,
such as the correlations between the ionized mass and 
nebular density with the radius, and those between particular
emission line ratios. 

In Sect.~\ref{sec_pnlf} we move to analyse the predicted  
synthetic PNLF, with the aim to discuss the expected behaviour  
of the bright cut-off, $M^*$. We will in particular focus  on
the nature of the objects responsible for the bright cut-off, analysing 
the dependence of $M^*$ on physical and technical factors.

Finally, Sects.~\ref{sec_discussion} and \ref{sec_conclusion} 
summarise the results of present models in terms of basic dependences
of the cut-off, and  
draw a few conclusions about the 
PNLF method, at the same time
highlighting the main points of discrepancy between our 
findings and the picture used so far to interpret the PNLF results.

%%%%%%%%%%%%%%%%%%%%%%%%%%%%%%%%%%%%%%%%%%%%%%%%%%%%%%
\section{Synthetic model for PN populations}
\label{sec_update}

\subsection{Overview of the PN model}
\label{sec_overview}

We follow the evolution of a single PN by using the synthetic
code from Paper~I. The main aspects of our PN
models are:
\begin{itemize}
\item The stellar evolution prior to the PN-phase 
is taken for a large grid of initial masses $\Mini$ 
and metallicities $Z$ from Girardi et al. (2000) for phases up to the
first thermal pulse, 
and from Marigo et al. (1999) and Marigo (2001) during the 
complete TP-AGB phase (the AGB-phase with thermal pulses
occuring). The TP-AGB evolution includes the main 
phase of mass loss, which later gives origin to the PN shell. 
We use the model predictions to 
follow the evolution in time and radius of the wind ejecta 
and their chemical composition. 
\item We use post-AGB evolutionary tracks from 
Vassiliadis \& Wood (1994) to model the evolution of the 
PN central star (CSPN). Such tracks are available
for both H- or He-burning regimes. A transition time, $\ttr$, 
is assumed between the end of the TP-AGB and the beginning of
the post-AGB track at $\Teff=10\,000$~K. The CSPN tracks allow 
the computation of quantities such as the flux of UV ionising 
photons and the velocity of the fast wind.
\item The PN shell is ionized by the UV photons and expands 
due to its initial ejection velocity. As a major improvement of 
synthetic models we include the following dynamical effects:
the shell shaping by the inner fast wind emitted from the hot CSPN
(following Volk \& Kwok 1985),
the successive engulfment of external layers expelled at early times
with lower expansion velocities, and the interaction of multiple
winds, including the effect of shell-thickening due to ionisation
(based on Kahn 1983).  
\end{itemize}

In the original model (Paper~I), the description of
ionisation is simplified. The electron temperature \Te\ in the nebula 
is assumed to be constant and close to $10\,000$~K 
for the entire H$^+$ region. Furthermore, the sizes of the
H$^+$, He$^+$, He$^{++}$ regions are derived from
simple considerations about the sizes of the different Str\"omgren
radii, i.e.\ the radii below which the corresponding 
ionising photons would be responsible for complete ionisation. 
Moreover, the size of the O$^{++}$ region, whence 
the \oiiiline\ line is emitted, is simply assumed to be equal to the
He$^+$ region; this latter assumption is based on the
similarity between the He$^+$ and O$^{++}$ ionisation potentials
(54.4 and 54.9 eV, respectively). 
Of course, these assumptions directly affect the predictions
for emission lines such as \heiiline\ and \oiiiline. 
Despite these simplifications, it is demonstrated in Paper~I that our
PN models can reproduce quite well the observed correlation between
the line  ratios \heiiline/\Hbeta\ and \oiiiline/\heiiline. 

%%%%%%%%%%%%%%%%%%% Figure %%%%%%%%%%%%%%%%%%
\begin{figure}
\resizebox{\hsize}{!}{\includegraphics{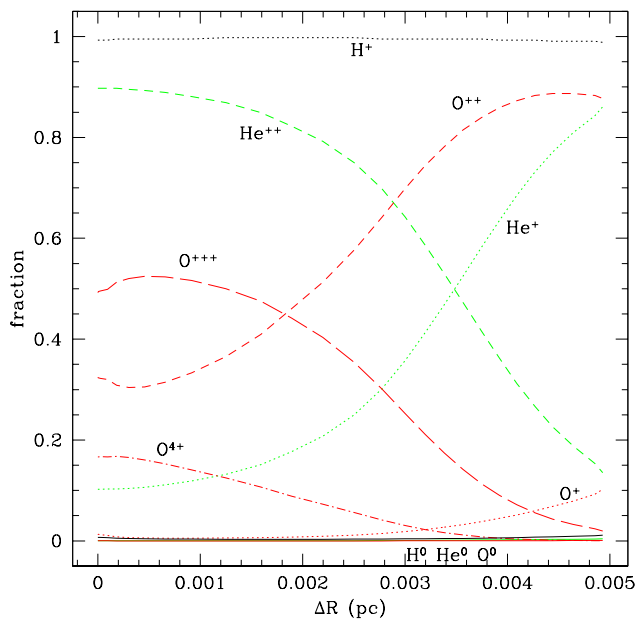}}
\caption{The ionisation fractions of H, He, and O ions,
as a function of radius inside the ionised PN shell. 
The ionisation structure has been 
computed with the aid of CLOUDY, as explained in 
Sect.~\protect\ref{ssec_cloudy}. This particular
PN model corresponds to the case $\Mini=2.0$~\Msun, $Z=0.008$, 
$\ttr=1\,000$~yr, H-burning track, at an age of $t=3\,000$~yr,
which is close to its maximum \oiiiline\ luminosity. Other 
relevant quantities are $M_{\rm CSPN}=0.685$~\Msun, 
$M_{\rm ion}=0.091$~\Msun, $R_{\rm in}=0.059$~pc, 
$\log\Teff=5.218$, and $\logL=3.867$.
}
\label{fig_structure}
\end{figure}

Nonetheless, these assumptions clearly are not adequate for an
accurate modelling of the absolute \oiiiline\ flux from PNe, 
which is our main target in the present paper because: 
(1) the flux of any collisionally-excited  
forbidden line is very sensitive to the electron 
temperature \Te, that therefore should not be assumed  constant,
but rather it needs to be consistently computed from the equations
of energy balance inside the nebular material; and 
(2) according to detailed photoionisation models, 
the O$^{++}$ ionised region is neither expected to have
sharp edges, nor is the O$^{++}$ fraction expected to  
follow so closely the He$^+$ one. These points are illustrated in 
Fig.~\ref{fig_structure}.

To overcome the limitations of Paper~I in this respect, the proper 
choice is to apply a complete photoionisation code to our 
evolutionary models.
This step is described in the following section.

\subsection{Incorporation of CLOUDY}
\label{ssec_cloudy}

We opt for the  photoionisation code CLOUDY 
(Ferland 2000, 2001), version 94,
which is a publicly-available, widely-used and well-tested 
tool for describing the structure of ionised regions and their
emission spectra. CLOUDY and other similar photoionisation codes 
have already been applied in the evolutionary modelling of PNe 
by a number of authors,
among them Volk (1992), Stasi\'nska (1989), 
Stasi\'nska \& Szczerba (2001)
Dopita \& Meatheringham (1990), Dopita et al. (1992),
Dopita \& Sutherland (2000), Phillips (2000).
However, in all those cases the treatment of the underlying stellar
population and of the nebula expansion was very crude, compared to our
method of Paper~I.

In order to couple CLOUDY to our evolutionary code, we proceed 
in the following way:
\benu
\item The former synthetic code (Paper~I) is run for
a guess electron temperature of $\Teo=10\,000$ K.
\label{step_ini}
\item At each time step, the radial structure (density, velocity,
chemical composition, etc.) of the ionised shell is taken from the
synthetic code and provided as input to CLOUDY.
\item CLOUDY is then used to solve for the energy balance, ionisation
structure, and emissivities as a function of radius. From the CLOUDY output,
we extract the most relevant information such as the volume-averaged
electron temperature \Te\ and the outcoming fluxes in some of the main
emission lines.
\label{step_last}
\eenu

To simulate the nebular structure given to CLOUDY a number of parameters
must be specified.  For the present work, 
we adopt an expanding-sphere geometry with covering 
factor equal to $1$ (see Ferland 2001 for details), 
inner and outer radii $R_{\rm S}$ and $R_{\rm ion}$, and
uniform hydrogen density $N({\rm H})$ (refer to  Paper~I
for the precise definitions of the quantities). 
All the 30 chemical elements allowed 
by CLOUDY are considered. For most elements, a scaled-solar 
abundance, corresponding to the adopted $Z$ value, 
is assumed.
The adopted fractions of H, He, C, N, O, Ne and Mg, on the contrary,
are the ones predicted from the evolution of
the PN progenitor, with all dredge-up events as well as hot-bottom
burning being followed in detail (see Marigo 2001).  
It should be remarked that the abundances of these elements 
in the main expanding 
PN shell may change with time, due to the 
accretion of successive outer layers ejected
from the progenitor star at previous stages of its 
nucleosynthetic evolution. Note that such a variable PN composition is
being considered for the first time in a theoretical model for the
PNLF. 

As for the central star, $L$ and \Teff\ values are converted 
into photon fluxes by using the following libraries of intrinsic
stellar spectra, already available in CLOUDY: 
ATLAS9 from Kurucz (1993) for $\Teff<50\,000$ K, 
and the PN nuclei non-LTE models from Rauch (1997) for 
$\Teff\ge50\,000$ K.

\subsection{Consistency and accuracy checks}

Initially, our intention was to iterate over the sequence of
operations \ref{step_ini} to \ref{step_last} outlined at the beginning
of Sect.~\ref{ssec_cloudy}, 
i.e. provide the output \Te\ from CLOUDY 
as a new \Teo\ guess to the synthetic PN model, and iterate steps 
\ref{step_ini} to \ref{step_last} until convergence to a 
single value of \Te\ had been reached.
In practice, it has become clear that this iteration is not necessary, for
the following reasons:

First, the output \Te\ values from CLOUDY turn out to be, in general,
of the same order as the input value of $10\,000$ K, 
especially during the phase in which the highest \Hbeta\ and 
\oiiiline\ luminosities are attained. Typical values are comprised
between $8\,000$ and a maximum value of $15\,000$ K. Lower \Te\ 
values are present only in the very initial phases of nebular 
ionisation (as the central star is heating up), and in very latest phases 
of recombination (as the central star fades on the white dwarf track).
These stages of low \Te\ present either very low nebular emissivities,
or very low surface brightnesses, and likely do not correspond to 
observable PNe.

%%%%%%%%%%%%%%%%%%% Figure %%%%%%%%%%%%%%%%%%
\begin{figure}
\resizebox{\hsize}{!}{\includegraphics{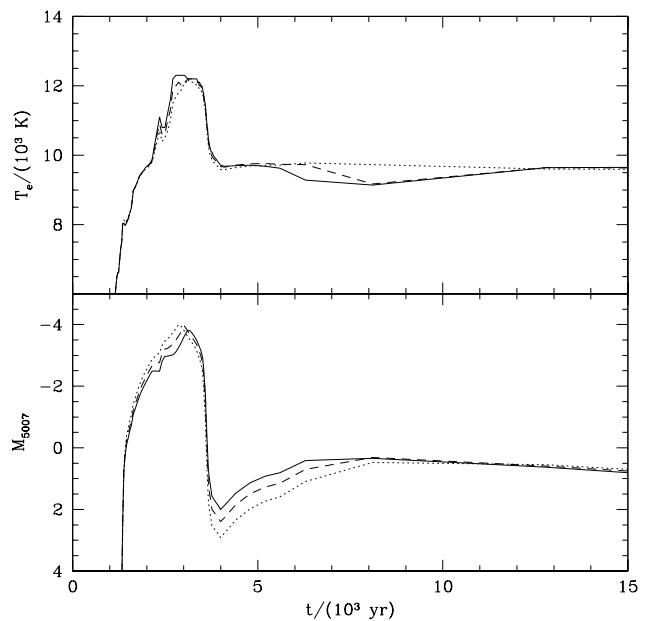}}
\caption{{\em Top panel:} The volume-averaged electron temperature, 
\Te, resulting from CLOUDY, as a function of time, for
a sequence of PN models computed for various values
of the input guess \Teo, namely 8\,000, 10\,000, and 12\,000~K
(dotted, dashed, and continuous lines, respectively). 
The models correspond to the case $\Mini=2.0$~\Msun, $Z=0.008$, 
$\ttr=1\,000$~yr, H-burning tracks.
{\em Bottom panel:} The resulting absolute magnitude in the 
\oiiiline\ line, \Moiii, as a function of time, for
the same PN models. \Moiii\ is only little affected by the value of
the input guess \Teo.}
\label{fig_tel_oiii}
\end{figure}
%%%%%%%%%%%%%%%%%%%%%%%%%%%%%%%%%%%%%

Second, the input \Teo\ provided to our synthetic code has a 
very small impact on the structure and evolution of the nebula,
and consequently hardly affects the
output \Te\ values from CLOUDY.
This is illustrated in Fig.~\ref{fig_tel_oiii} which presents the
output \Te\ from CLOUDY for a few models, corresponding to 
the same PN progenitor star but adopting three different values 
the input temperature \Teo, i.e.  $8\,000$, $10\,000$, and  
and $12\,000$ K.
Evidently, the output \Te\ values differ just 
little between the various cases. For the main phase of PN 
evolution, in the age interval between 2\,000 and 4\,000~yr
(where the \oiiiline\ luminosity attains the highest values), 
the \Te\ values differ by less than 4\%. The differences are caused by the 
modest changes in the density -- and hence radial thickness -- 
of the nebular shell, for which the input \Teo\ plays a minor role.

For this same set of models, Fig.~\ref{fig_tel_oiii} also presents 
the resulting absolute magnitude in the \oiiiline\footnote{\Moiii\ is 
computed from Eq.~(\protect\ref{eq_moiii}), assuming a PN
distance of 10 pc.}, \Moiii, as a function of time.
Again, the trial input \Teo\ has just a modest impact on the 
line flux calculated by CLOUDY. The largest differences are of 
about $0.3$~mag for the models illustrated in  Fig.~\ref{fig_tel_oiii};
we find that at larger stellar masses, that produce the brightest PNe, 
the differences are even lower. From tests we have performed,
we expect that on average, by using a constant 
\Teo\ of $10\,000$ K, the typical PN models will have
\Moiii\ magnitudes as accurate as to about $0.15$~mag. 

Such an accuracy may not seem good enough for the study of a
potential standard candle. However, we should recall that 
this theoretical indetermination of 
$M^*$ is comparable to the global uncertainty
%\footnote{The observational 
%uncertainty on $m^*$ includes contributions related to the i) fitting procedure
%($\approx 0.10$ mag), ii) definition of the empirical 
%PNLF ($\approx 0.05$ mag), iii) photometric zero points  ($\approx 0.05$ mag),
%and iv) filter response calibration  ($\approx 0.04$ mag).}
affecting the observational derivation of $m^*$ (see Jacoby et al. 1990).
Moreover, we expect that the errors in our predicted \Moiii\ 
will be mostly systematic, whereas our study intends 
to explore mainly the differential effects -- i.e. 
the changes in $M^*$ between different galaxies. 
We also remark that, although the absolute values
of $M^*$ is of high interest, an accurate theoretical
calibration of its value is still out of reach, as we 
will demonstrate in this paper. 

\subsection{Summary of parameters}
\label{sec_parameters}

The low sensitivity of our results to the input guess \Teo, 
effectively  cancels this quantity from our list of input 
parameters. The adjustable parameters that remain in our models 
are:
\bite
\item the initial mass \Mini\ and metallicity $Z$;
\item the transition time \ttr;
\item the choice of either H- or He-burning post-AGB track.
\eite
Other important quantities like the initial He content, and the
CSPN mass, are defined by the relationships already incorporated 
in the synthetic evolution models, such as the helium-to-metal 
enrichment ratio 
(see Girardi et al.~2000), and the initial--final mass relation 
(see Marigo 2001), which has a dependence on metallicity.

The typical values for \ttr, and the type of post-AGB 
evolution, have a great impact on the evolution 
of single PN. For a given mass and metallicity, they 
completely settle the evolutionary timescale of the central 
star, and hence the duration of the PN phase.
It is worth recalling the present uncertainties in these
key ingredients of our models:

The transition time \ttr\ is very badly predicted by theory.
Its value depends on several details of the final phase of
mass loss after the termination of the AGB-phase, when the star is
getting hotter. 
Main factors driving \ttr\ are the
velocity by which the residual stellar envelope is peeled off
from the star, and the velocity by which the same envelope is
burnt from below by nuclear reactions.  
Educated guesses about these processes can be made
(see e.g.\ the discussion in Stanghellini \& Renzini 2001), but
lead to a general picture of great uncertainty, in which the
only firm conclusions are that \ttr\ should generally decrease
with stellar mass, and have values of the order of hundreds to 
thousands of years. Alternatively, \ttr\ values can be
evaluated from evolutionary codes including
shell-burning and reasonable prescriptions for mass-loss
(e.g. Wood \& Faulkner 1986).
{From} such models, Vassiliadis \& Wood (1994)
find higher \ttr\ values between 3\,000 and 20\,000~yr, but
with no obvious correlation with mass. 
Moreover, the derived \ttr\ depends critically
on the particular phase of the pulse cycle at which the star 
leaves the AGB. Therefore, \ttr\ is very uncertain,
and even for the optimistic case in which the mass-loss rates were 
known in detail, a large star-to-star variation of \ttr\ might 
be expected.

In our models, we decided to compute PN grids for single values
of constant \ttr. This approach is useful for calibrating
the typical values of \ttr, which is otherwise
essentially unknown. In Paper~I, we found that data about Galactic
PNe, as for example the relation between mass and radius of the
ionized matter, can be reproduced well with shorter transition times
of 1000~years or below.

The type of post-AGB evolution is also determined, mainly, by the
phase of the pulse cycle in which the star leaves the AGB
(Iben 1984; Wood \& Faulkner 1986; Vassiliadis \& Wood 1994):
if it occurs during the quiescent phases between thermal pulses,
a classical H-burning track is followed, otherwise the central star
starts the post-AGB as a He-burning object. 
In principle, both H- and He-burning cases are to be expected
in samples of PNe. The main challenge of theory, in this case,
is to constrain the expected ratio of H- and He-burning stars.

In our models, we compute PN grids for either H- or
He-burning tracks. The resulting synthetic samples can then be 
mixed in varying proportions, before they are compared with data.

%\subsection{Grids of PN models}

For the present study extensive 
grids of PN models are calculated with the following input parameters:
\bite
\item $Z=0.004$, $0.008$, $0.019$, which correspond rougly to
the SMC, LMC, and solar composition;
\item $\ttr=500$, and $1\,000$~yr; 
\item both H- and He-burning tracks. 
\eite
Each grid corresponds to at least 20 tracks computed
for stellar masses between $\Mini=0.6$ and $5$~\Msun. 
All models adopt the $f(L)$ prescription for the terminal 
wind velocity (Paper~I). Additional sets with different
\ttr\ values -- namely 800 and 2500 yr -- are calculated
for the $Z=0.019$, H-burning case.

%%%%%%%%%%%%%%%%%%%%%%%%%%%%%%%%%%%%%%%%%%%%%%%%%%%%%%%%

\section{Properties of synthetic PN samples}
\label{sec_samples}

\subsection{Modelling populations of PNe}

In order to simulate populations of PNe in external galaxies,
we employ a simple population synthesis approach. Since
the PN lifetimes, $\Delta t^{\rm PN}$, 
are much shorter than the main sequence
lifetimes of their progenitors, $\tau^{\rm H}$, 
the initial mass distribution of PNe -- the number of PNe from stars
with initial mass $\Mini$ -- is fairly well 
described by:
\beq
N(\Mini) \propto \phi(\Mini)\,\psi(t-\tau^{\rm H})\,\Delta t^{\rm PN} \,\,,
\label{eq_massdist}
\eeq
where $\phi(\Mini)$ is the initial mass function (IMF), and $\psi(t)$ is
the star formation rate (SFR, in mass per unit time) at the galactic
age $t$. 

In the present work, we simply assume a Salpeter (1955) IMF, 
$\phi(M)\propto\Mini^{-2.35}$. The stellar quantities, $\tau^{\rm H}(\Mini,Z)$
and $\Delta t^{\rm PN}(\Mini,Z)$, are interpolated from the existing
grid of models to any given value of initial mass $\Mini$ and
metallicity $Z$. Given an age $t$, the stellar metallicity 
is selected from a given age--metallicity relation (AMR) $Z(t)$. 

By assuming a given star formation history 
$\psi(t)$ and $Z(t)$, Eq.~(\ref{eq_massdist}) is used to construct the
probability distribution function of initial masses. 
Then, any synthetic sample of PNe is randomly generated with the
aid of a simple Monte Carlo method, provided that the above distribution
is complied with. 
Hence, for each star, a random age is selected in the
interval $[0,\Delta t^{\rm PN}]$. Finally, all stellar parameters
(nebular masses, radii, velocities, emission fluxes, and the
properties of central stars) are obtained from
interpolation in the existing PN model grids.

It should be recalled that the resulting synthetic samples strictly 
obey the basic theories 
of stellar evolution and population synthesis of galaxies, and therefore they 
present some of the intrinsic selection effects and biases that 
are inherent to these theories. For instance, tracks with 
higher PN-phase lifetimes are naturally more represented
in these samples.
In order to simulate observed samples, additional selection 
criteria should be imposed, such as e.g.\ a given limit 
in \oiiiline\ magnitude and/or surface brightness.
In the present work, such selection criteria are assumed
a posteriori, when required for the comparison with empirical data.

For each one of PN model grids, we simulate synthetic 
PN samples until a number of $10^4$ PNe brighter than 
$\Moiii=0$ is reached. 
This guarantees an adequate sampling of the brightest 
part of the PNLF.

%%%%%%%%%%%%%%%%%%%%%%%%%%%%%%%%%%%%%%%%%%%%%%%%%%%%%%%%%%%%%%%%%%
\subsection{Comparison with empirical data}
\label{sec_comparison}

\begin{figure}
\resizebox{\hsize}{!}{\includegraphics{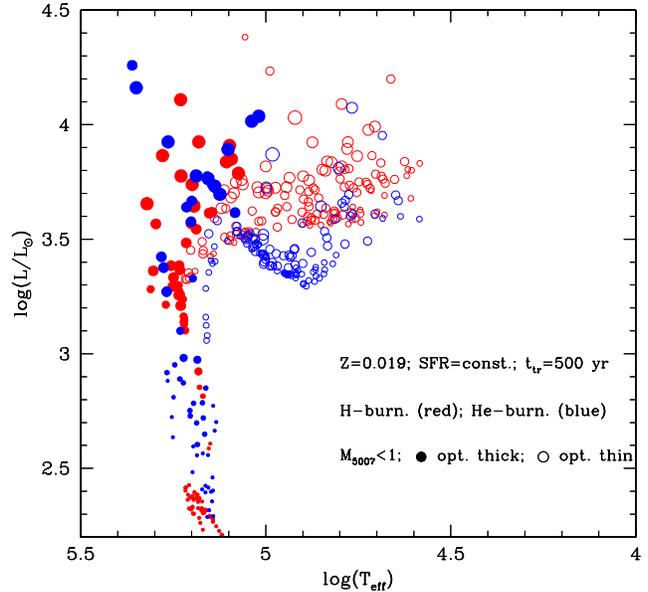}}
\caption{Location of PN central stars in the HR diagram.
The simulated sample corresponds to PNe with $\Moiii < 1$ out of
a total population of 500 objects, assuming the following set of
input parameters: constant SFR, constant metallicity $Z=0.019$, 
transition time $\ttr=500$ yr, H- or He-burning post-AGB evolutionary tracks.
Filled and empty circles denote PNe that are optically thick and 
optically thin to the hydrogen ionising photons, respectively.
Circles are larger for brighter \Moiii}
\label{fig_hr}
\end{figure}

\begin{figure}
\resizebox{\hsize}{!}{\includegraphics{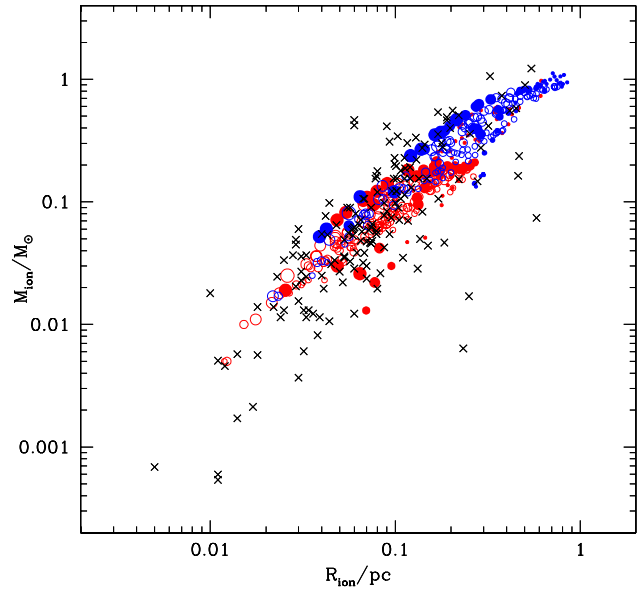}}
\caption{Ionised mass--radius relation for Galactic PNe.
Observed data (crosses) are taken from Zhang (1995) and 
Boffi \& Stanghellini (1994). The synthetic sample (circles) is
the same one as in Fig.~{\protect\ref{fig_hr}}}
\label{fig_masrad}
\end{figure}

\begin{figure}
\resizebox{\hsize}{!}{\includegraphics{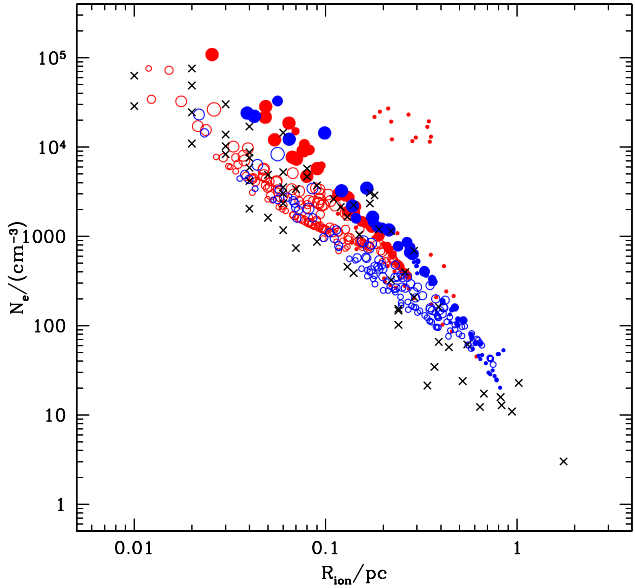}}
\caption{Electron density--radius relation for Galactic PNe.
Observed data (crosses) are taken from Phillips (1998). 
The synthetic sample (circles) is
the same one as in Fig.~{\protect\ref{fig_hr}}}
\label{fig_nerad}
\end{figure}

\begin{figure}
\resizebox{\hsize}{!}{\includegraphics{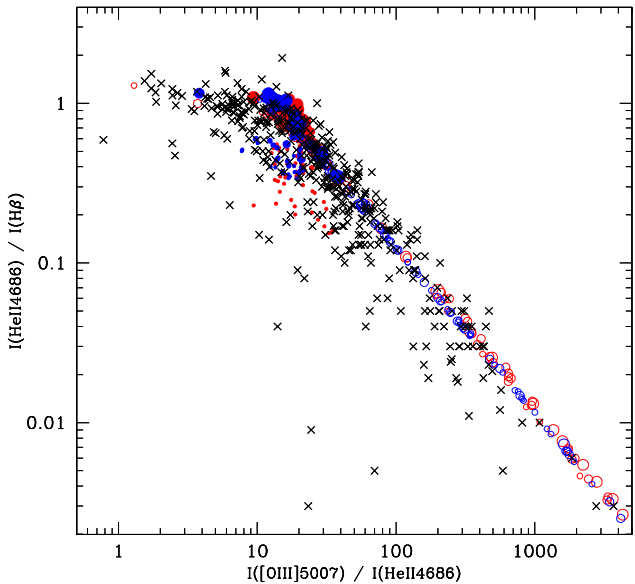}}
\caption{Anticorrelation between \heiiline\ and \oiiiline\ 
line intensities. Observed data (crosses) are taken from the Strasbourg-ESO
Catalogue of Galactic Planetary Nebulae (Acker et al. 1992).
The synthetic sample (circles) is the same one as 
in Fig.~{\protect\ref{fig_hr}}}
\label{fig_lineratio}
\end{figure}

Figures \ref{fig_hr}--\ref{fig_lineratio} illustrate the predicted
properties of synthetic samples of PNe and their central stars 
for exemplifying choices of the input parameters (namely: constant 
metallicity $Z=0.019$; $\ttr = 500$ yr; H/He-burning tracks; constant
SFR) and selection criteria (specified in the captions). 
Similar plots can be obtained for different input parameters. 

%(i.e.  we consider only those PNe 
%with $\Moiii < 1$ out of a total population of 500 objects).
 
Let us start with Fig.~\ref{fig_hr}  that shows 
the predicted distribution of the PN central
stars in the HR diagram, as a function of the \oiiiline\ luminosity
(larger circles correspond to more luminous PNe).
On average, central stars with higher luminosity and larger
effective temperature produce brighter PNe 
in the \oiiiline\ line. We see that PNe -- emitting in the  \oiiiline\ line --
should be optically-thin to the H-ionising flux as long as their central stars
are evolving along the horizontal part of their post-AGB tracks; later they 
may become optically-thick as the PN nuclei start fading.
Moreover, we notice that  He-burning central stars of 
optically-thick PNe tend to concentrate towards 
lower luminosities compared to H-burning objects.

%%%%%%%%%%%%%%%%%%% Figures %%%%%%%%%%%%%%%%%%

\begin{figure}
\resizebox{\hsize}{!}{\includegraphics{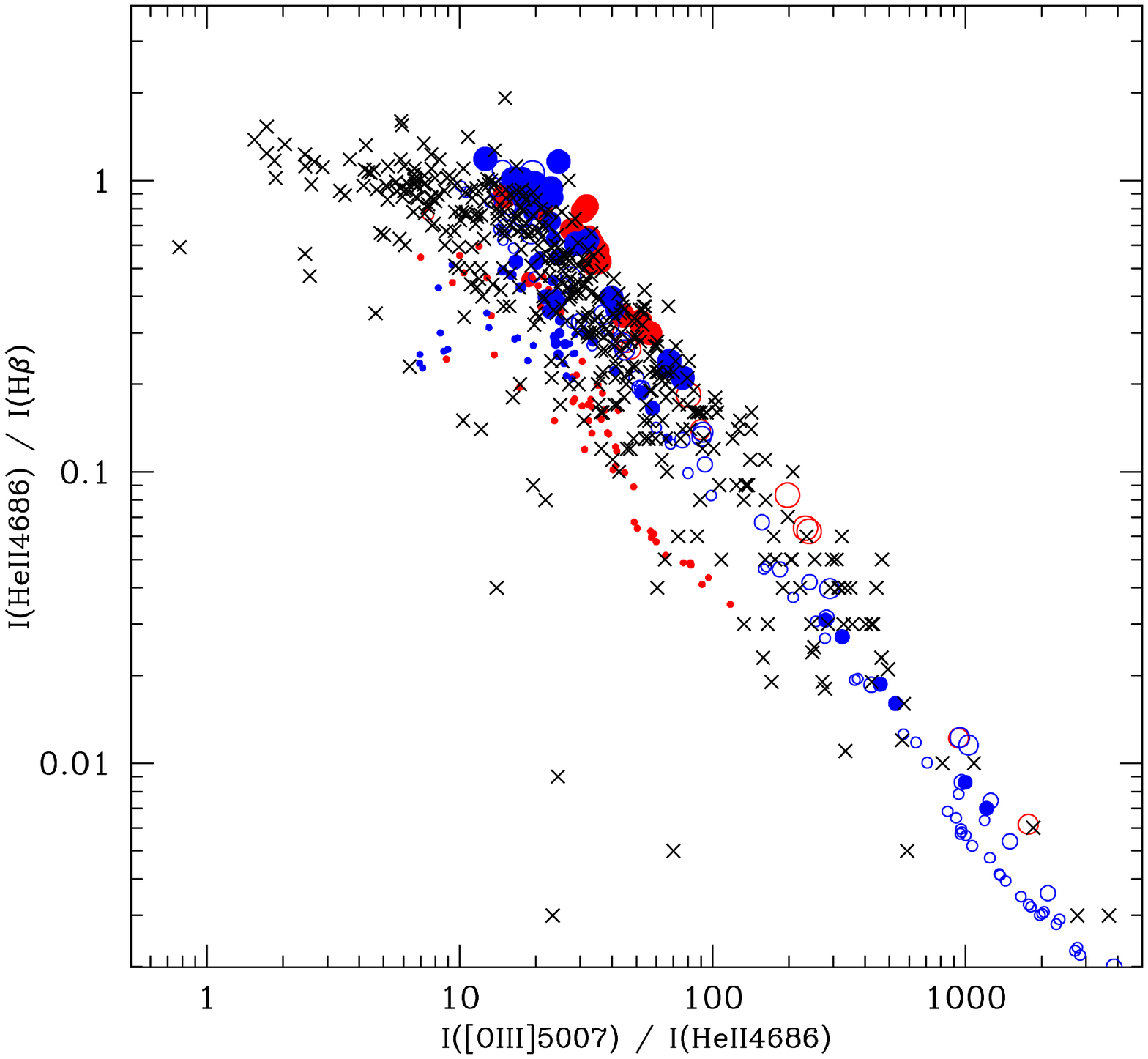}}
\caption{Anticorrelation between \heiiline\ and \oiiiline\ 
lines. The same as Fig.~{\protect\ref{fig_lineratio}}, but for
$Z=0.008$ }
\label{fig_lineratioz008}
\end{figure}

\begin{figure}
\resizebox{\hsize}{!}{\includegraphics{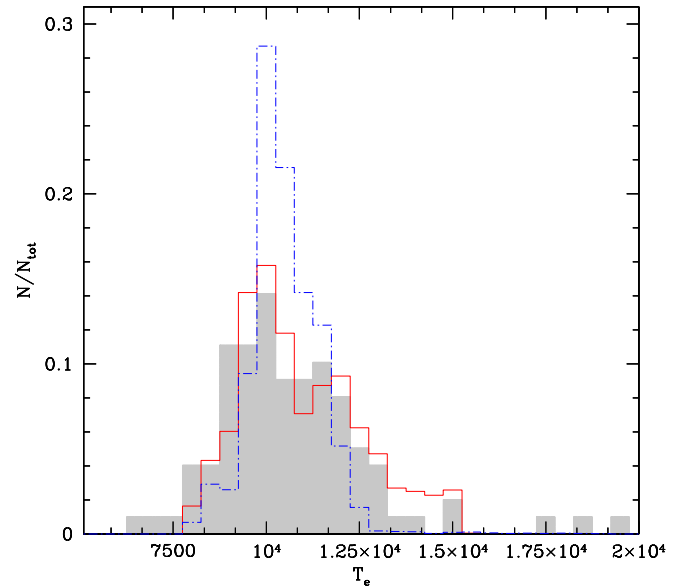}}
\caption{Distributions of electron temperatures in PNe. 
Shaded histogram:
observed distribution of $\Te [\mbox{O\,{\sc iii}}]$ 
for a sample of 99 Galactic PNe taken from 
McKenna et al.\ (1996). Empty histograms:
predicted distributions derived from synthetic samples 
with parameters $Z=0.019$, $\ttr=500$ yr, H-burning tracks 
(continuous line) and He-burning tracks (dot-dash line)
  }
\label{fig_tedist}
\end{figure}

\begin{figure}
\resizebox{\hsize}{!}{\includegraphics{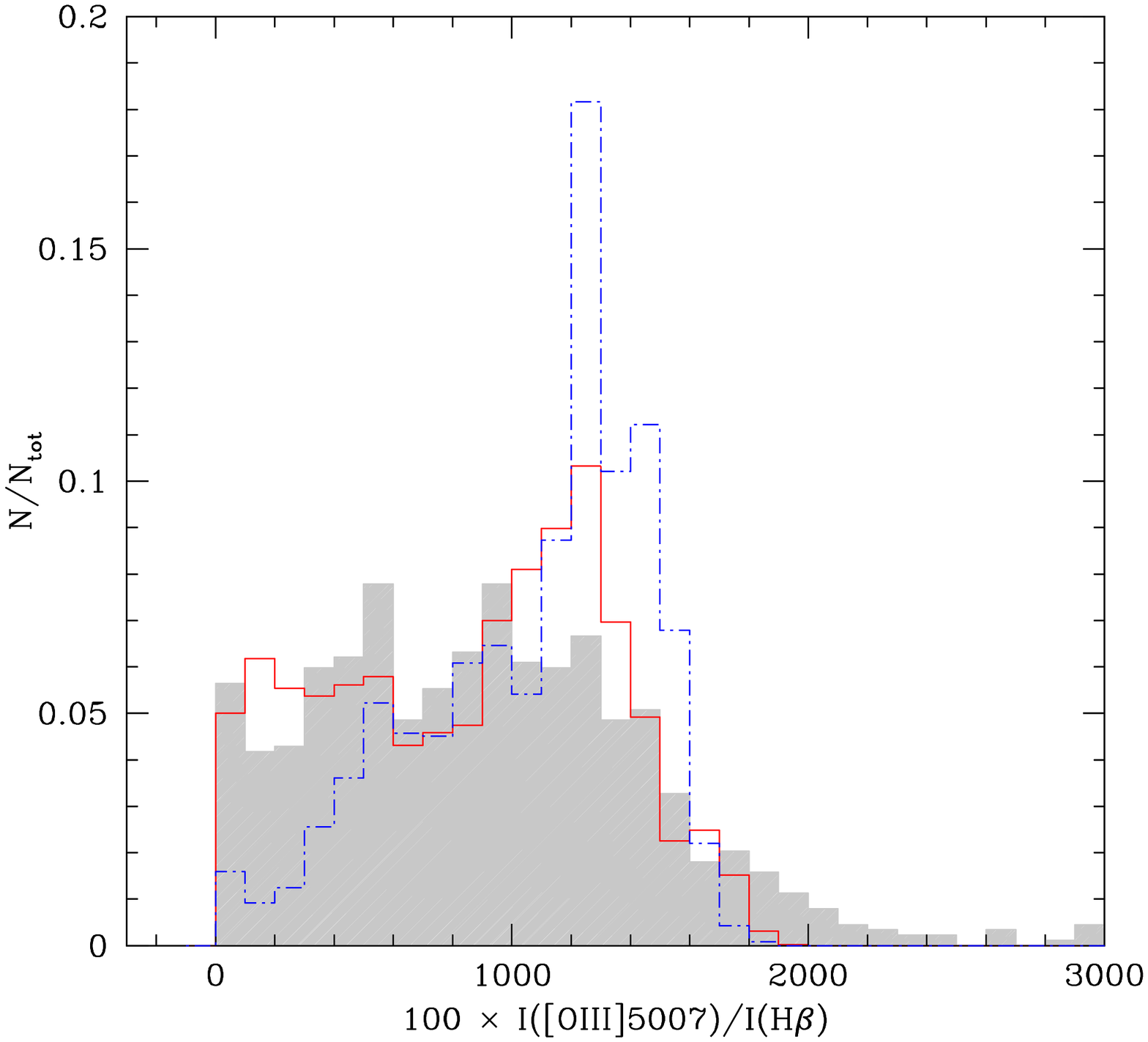}}
\caption{Distributions of \oiiiline\ intensity relative to
\Hbeta, on the scale $I(\Hbeta)=100$. Shaded  histogram:
observed distribution for 886 Galactic PNe taken from the 
ESO-Strasbourg Catalogue (Acker et al.\ 1992). Continuous-line histogram:
predicted distribution derived from synthetic samples 
with parameters $Z=0.019$, $\ttr=500$ yr, H-burning tracks, re-scaled
to the total number of observed PNe. Dot-dashed line  histogram:
the same but for He-burning tracks. We remark that the 
corresponding distributions for $\ttr=1000$ yr change little 
with respect to those already plotted}
\label{fig_oiiidist}
\end{figure}

Figures \ref{fig_masrad}--\ref{fig_lineratio} 
display how synthetic samples of PNe 
distribute over a few relevant
diagrams, where observed PNe are found to follow clear correlations
between the involved nebular parameters, namely: 
the ionised mass, the nebular radius, the electron density and 
specific intensity line ratios.
For a complete discussion of the diagrams we refer to Paper~I, where
the empirical data are compared to predictions of  
individual parameters as a function of the 
PN age for a representative set 
of PN models. In our present study, the comparison is instead  
performed in terms of statistical properties, with the aid of synthetic
populations of PNe. 

A few remarks should be made at this point.
First, we see that the synthetic samples  recover 
the observed data quite well, which confirms the good agreement already
obtained in Paper~I.

For the ionised mass-radius and electron density -- radius relationships 
(Figs.~\ref{fig_masrad} and  \ref{fig_nerad}) 
we have the interesting result that, 
for the given set of adopted parameters, the observed excursion 
in radius is accounted for by including both H- and He-burning
tracks, the latter group mainly contributing at larger radii.
This is explained by the larger evolutionary timescales of the
He-burning stars, so that more expanded nebulae are expected at
given ionised mass.
We also note  that the observed points at lower masses in 
Figs.~\ref{fig_masrad} could be
recovered by widening the selection criteria in our simulated sample,
i.e. including also PNe that are fainter than $\Moiii=1$ (which is our
present upper limit in magnitude).

The empirical anticorrelation between the \heiiline\ and \oiiiline\ line 
intensities, displayed in Fig.~\ref{fig_lineratio}, is well reproduced by
the synthetic sample for metallicity $Z=0.019$, 
which however follows a narrower relationship than
observed. A larger dispersion is obtained by a simulated PN sample
with the same parameters but for a lower metallicity, i.e. $Z=0.008$
(see Fig.~\ref{fig_lineratioz008}). 
Intepreting  this result in terms of the oxygen abundance $X($O$)$, we may 
derive the indication that Galactic PNe derive from stellar
progenitors with lower $X($O$)$ than assumed in the solar-metallicity models
(Anders \& Grevesse 1989). To this respect,  it is worth recalling that, 
according to the recent determination by Allende Prieto et al. (2001),
 the Sun's  oxygen abundance is lower by about  0.2 dex than 
previous estimates (e.g. Anders \& Grevesse 1989), that 
are actually  adopted in our solar-metallicity models.
%Finally, with respect to Fig.~\ref{fig_lineratioz008} it should 
%also be remarked that
%in this case the sample is extended to include PNe with $\Moiii < 5$, 
%which allows to reach 
%the data towards higher I(\heiiline)/I(\Hbeta). 

Finally, Figs.~\ref{fig_tedist} and ~\ref{fig_oiiidist} show the comparison
between observed and predicted distributions of two relevant quantities, 
namely: the electron temperature and the line ratio $I(\oiiiline)/I(\Hbeta)$,
on the scale $I(\Hbeta)=100$.
The empirical data refer to samples of Galactic PNe.

As already mentioned, the electron temperature is a critical quantity 
when dealing with the intensities of collisionally-excited lines, such 
as \oiiiline. It follows that the distribution of \Te\ represents 
an important observable to test our PN models, particularly in relation
with the \oiiiline\ PNLF. We see in Fig.~\ref{fig_tedist} that the comparison
turns out successfully as models are able to recover the location
of the distribution peak, centered at around $\Te=10\,000$ K.
We also notice that the the H-burning set of models 
appears to describe well all the main features of the observed \Te\ 
histogram, the He-burning set populating  a narrower \Te\ range 
than observed.    
    
The distribution of the \oiiiline\ intensity relative to $\Hbeta$
also represents a relevant constraint to model predictions. 
It is worth remarking that a satisfactory reproduction of the
observed data is naturally obtained with our PN models -- mostly
the H-burning set --  without introducing any additional assumption
or free  parameter. Again, this sets another important improvement
on previous studies of the PNLF (e.g. M\'endez \& Soffner 1997), in which
the  reproduction of the $I(\oiiiline)$ distribution -- 
without the use of a photoionisation code -- is derived analytically and it 
requires a number of calibration steps, i.e. 
the  $I(\oiiiline)$ values are adjusted during the simulations 
in a fashion depending on the mass of the central star 
and its position in the H-R diagram.  

%%%%%%%%%%%%%%%%%%%%%%%%%%%%%%%%%%%%%%%%%%%%%%%%%%%%%%%%

\section{Properties of synthetic PNLFs}
\label{sec_pnlf}

In this section, we will discuss the dependecy of the
PNLF cut-off on several parameters. With the aid of our 
models, we start answering the
most basic question, namely which stars produce the PNe populating
the cut-off (Sect. \ref{sec_whatcutoff}). 
Then, we analyse the intrinsic dependencies on two poorly known
model parameters, namely the dominant nuclear burning regime  
of post-AGB central stars (Sect. \ref{sec_HHetracks}) and the transition time 
(Sect. \ref{sec_ttr}). Finally, we move to investigate the effect of 
two genuine parameters of stellar populations -- that are likely to vary 
from galaxy to galaxy --, namely age and metallicity  (or rather the 
oxygen abundance) of PN progenitors 
(Sects. \ref{sec_ageburst} and \ref{sec_metal}, 
respectively). 
%Finally, we discuss how our
%results would change if other effects like other progenitor models 
%and self-extinction were considered in the simulations 
%(Sect. \ref{sec_otherdepend}). 

%%%%%%%%%%%%%%%%%%%%%%%%%%%%%%%%%%%%%%%%%%%%%%%%%%%%%%%%
\subsection{Which PNe form the cut-off\,?}
\label{sec_whatcutoff}

This simple and basic question has received different answers by
different authors. Just to mention two of them, M\'endez et al. (1993)
conclude that the cut-off PNe are optically-thin objects, whereas
Dopita et al.\ (1992) derive that some of 
the brightest LMC PNe are optically thick. 
M\'endez et al. (1993), M\'endez \& Soffner (1997), and 
Dopita et al.\ (1992) conclude that the cut-off PNe are young objects, 
at least in star-forming galaxies like the LMC (i.e. $\sim 0.8$ Gyr), whereas
Ciardullo \& Jacoby (1999) argue that, due to circumstellar extinction 
 mainly affecting the most massive AGB stellar progenitors, 
the brightest observed PNe in galaxies have always higher ages 
(i.e. $>1$ Gyr). 
In our models, we attempt to answer this question just by analysing  
the brightest PN models in our available grids, without 
making any additional assumption about their properties.

\begin{figure*}
\includegraphics[width=0.8\textwidth]{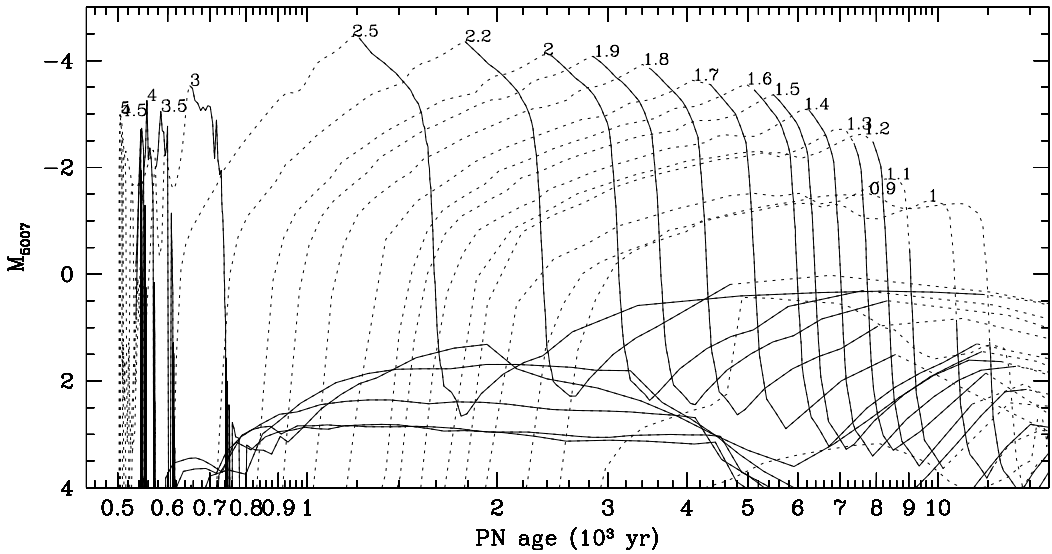}
\caption{Plot of \Moiii\ versus PN age for the tracks in our
grid of $Z=0.008$, $\ttr=500$ yr, H-burning tracks. 
It can be noticed that the most luminous PNe are those
of mass close to 2.5 \Msun. Optically thick nebulae are 
marked with continuous lines, optically thin with dotted ones.}
\label{fig_whatcutoff}
\end{figure*}

\begin{figure*}
\begin{minipage}{0.77\textwidth}
\includegraphics[width=\textwidth]{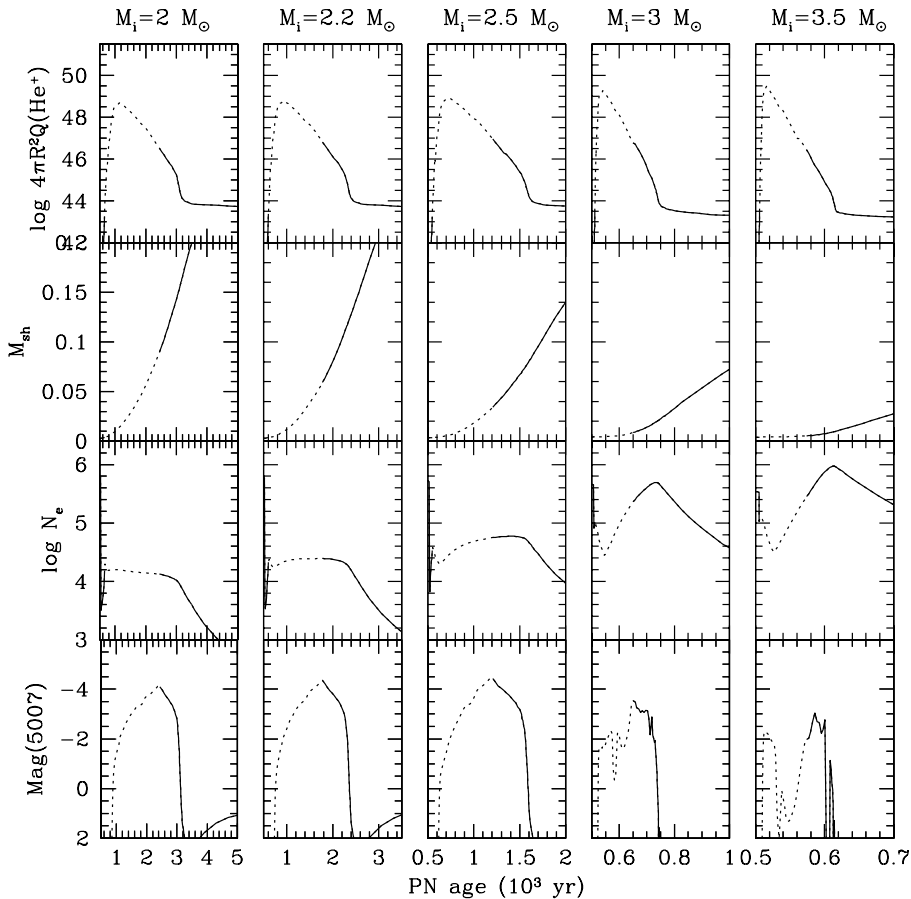}
\end{minipage}
\hfill
\begin{minipage}{0.21\textwidth}
\caption{Several model quantities plotted as a function
of PN age, for initial masses between 2 and 3.5 \Msun,
in our grid of $Z=0.008$, $\ttr=500$ yr, H-burning tracks. 
The age interval covers the complete period in which the object
is seen as a bright PN in \Moiii. Optically-thick nebulae are 
marked with continuous line, those optically-thin with dotted line.
The 2.5-\Msun\ track, responsible for the PNLF cut-off,
is the most massive one for which the periods of high 
photoionisation rate $4\pi R^2Q({\rm He^+})$
partially coincide with a significant 
shell mass $M_{\rm sh}$, and with an electron density 
$N_{\rm e}$ below the critical value for collisional 
de-excitation of the \oiiiline\ line. For lower-mass stars, 
the shell masses become higher, but the photoionisation rates
much lower
}
\label{fig_paneltracks}
\end{minipage}
\end{figure*}

We start by considering 
Fig.~\ref{fig_whatcutoff} that displays the run of \Moiii\ versus age, 
for the grid of $Z=0.008$, $\ttr=500$ yr, H-burning tracks. 
This grid was chosen mainly because it presents, for the
case of continuous SFR, a cut-off magnitude $M^*=-4.63$ 
(see Fig.~\ref{fig_PNLFhhe}, continuous-line histogram), 
very close to the observationally-calibrated value of 
Ciardullo et al.\ (2002). From Fig.~\ref{fig_whatcutoff} it 
turns out that this 
cut-off $M^*$ coincides with the maximum \Moiii\ reached by the
2.5~\Msun\ track, which defines the characteristic
progenitor stellar mass of cut-off PNe. Similar results are obtained 
for each grid of tracks; what varies from case to case
is just this characteristic mass, which tends to be slightly 
lower for higher thansition times and for He-burning 
tracks. Anyway, in all cases this characteristic mass is 
comprised between 2 and 3 \Msun.

What determines the \oiiiline\ maximum luminosity of 
the individual tracks\,? For most of them, this
maximum occurs when the PN switches from
optically thin to optically thick, 
i.e.\ when \muH\ changes  from 
$<1$ to $=1$. In fact, at that point the central star
starts dimming towards its WD cooling track,
no longer providing enough photons to completely ionize the 
nebula, so that the Str\"omgren sphere suddenly starts decreasing. 
At the same time the O$^{++}$ region also starts receeding in radius, 
and hence to the \oiiiline\ luminosity decreases. 

From these considerations it follows that the 
PNe observed close to their maximum \oiiiline\ luminosity,
can be either optically thin or thick, depending on whether 
they are evolving along the ascending or descending part 
of their tracks in  Fig.~\ref{fig_whatcutoff}.
This implies that both thin and thick nebulae
should be present in the PNLF cut-off, the frequency of 
optically-thin objects likely being higher given the faster evolution 
evolution of \Moiii\ in the descending branch.
In this detail, our results differ
from the assumptions/conclusions of both M\'endez et al. (1993) and 
Dopita et al. (1992).

At this point it is natural to wonder: 
Why is the absolute \Moiii\ minimum value attained just by the 
2.5 \Msun\ tracks\,? The answer to this question is not trivial,
since a lot of factors are envolved. 
Our proposed explanation is based on the temporal behaviour of 
three quantities, that are plotted in Fig.~\ref{fig_paneltracks} 
as a function 
of the PN age for a few models with CSPN masses around 2.5 \Msun. 
They are: 

I) The ionisation rate for O$^{++}$, that expresses the number of 
available ionising photons emitted by the central star per unit time. 
This rate is given by  $4\pi R^2Q({\rm O^{++}})$, where $R$ is the stellar
radius and $Q({\rm O^{++}})$ is the ionising flux. We recall that,
due to the similarity between their ionisation potentials,
O$^{++}$ and He$^+$ have similar ionisation fluxes $Q$ (see Paper~I). 
Therefore the behaviour of the ionisation rate for O$^{++}$ mimics that
of He$^+$, which is displayed in Fig.~\ref{fig_paneltracks}.
We see that larger ionisation rates correspond to more massive stellar
progenitors, as they produce more massive and luminous central stars.
Therefore,  models with progenitor masses of 3.0 and 3.5 \Msun\  
could be potentially brighter in \oiiiline\ than the 2.5 \Msun\ model. 

II) The maximum nebular 
mass available to the O$^{++}$ ionising photons. This 
is given, at each instant, by the mass of the shocked shell, $M_{\rm sh}$,
which is steadily increasing with time. It is clear that, in principle,  
a higher \oiiiline\ luminosity is favoured at larger $M_{\rm sh}$.

III) The electron density, $N_{\rm e}$. In fact, it is 
well-known that electrons do play the major
role in populating by collisions the upper level of \oiiiline\, so that
the higher $N_{\rm e}$ the stronger the spectral line. 
This holds until a critical density  $N_{\rm e}^{\rm crit}$ is reached, above
which the collisional de-excitation becomes important and the
the intensity of \oiiiline\  starts to weaken.
Moreover, we see that $N_{\rm e}$ steadily decreases with time.

\begin{figure}
\includegraphics[width=\columnwidth]{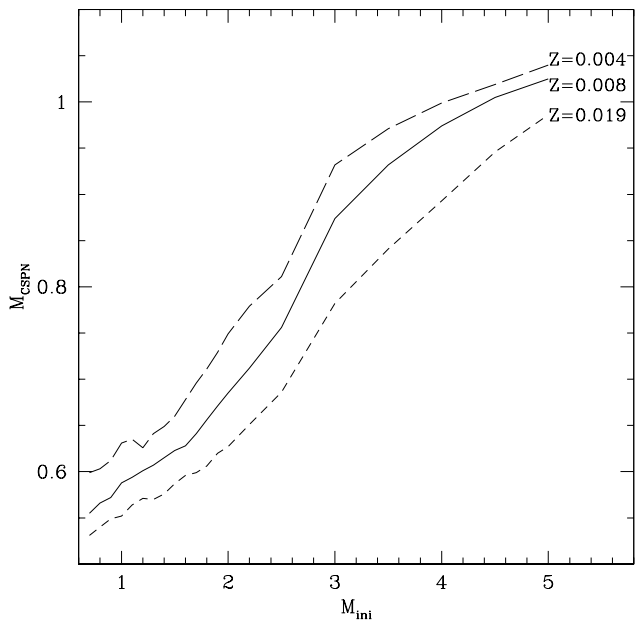}
\caption{Theoretical IFMR adopted in 
the present calculations. 
Notice that at given initial masses at decreasing metallicity 
models predict higher final masses, 
hence more massive central stars of PNe.}
\label{fig_ifmr}
\end{figure}

The answer to the initial question rests on the behaviour of all 
above quantities, whose evolution is controlled by the timescales 
of both central star and nebular dynamics.
From Fig.~\ref{fig_paneltracks} it turns out that during the initial
PN evolution, i.e. including the stages that bracket 
the ionisation rate's maximum, only  a fraction of the  hydrogen ionising 
photons  are effectively absorbed by the nebula that will be optically thin 
($\muH<1$), while a large fraction will contain O$^{++}$.
The comparison between the 
panels for different masses in 
Fig. \ref{fig_paneltracks} then reveals the following.

On one hand, for the most massive tracks ($\ga3$ \Msun) the evolutionary rate
of the central star is so rapid that, during the stages of significant
ionisation rate, the shocked shell has engulfed a relatively little amount  
of mass. It means that, despite the large availability 
of energetic photons to ionise O$^{++}$, this is effectively reduced by 
the little availability of nebular matter.  
Additionally, we notice that at these ages a further weakening of
\oiiiline\ emission may be produced as  
the nebular densities approach, or even 
overcome,  the critical limit for collisional de-excitation of
the \oiiiline\, upper level, i.e. $\sim 4.9\times10^5\,{\rm cm}^{-3}$.
All these factors concur to prevent PN models, having progenitor stars with 
$M > 2.5 M_{\odot}$, from becoming brighter in  \oiiiline\ than
those produced by stars with $M \sim 2.5 M_{\odot}$.  

\begin{figure*}
\begin{minipage}{0.77\textwidth}
\includegraphics[width=\textwidth]{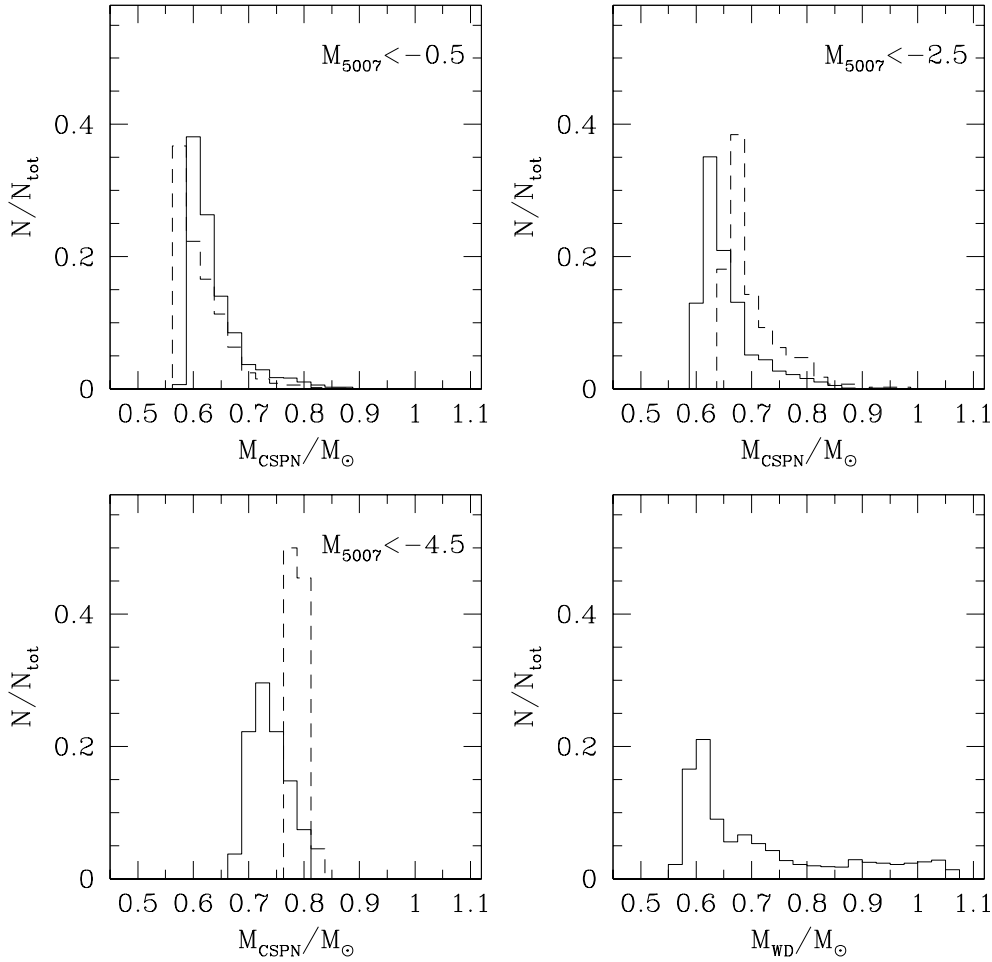}
\end{minipage}
\hfill
\begin{minipage}{0.21\textwidth}
\caption{Theoretical mass distributions for initial metallicity
$Z=0.008$. The left and top-right panels refer to central stars of PNe
more luminous than a limit magnitude \Moiii\ as indicated, with continuous 
and dashed lines corresponding  to H- and He-burning stars, 
respectively. For the sake of comparison, the predicted mass distribution
of white dwarfs is shown in the  bottom-right panel
}
\label{fig_mcsp}
\end{minipage}
\end{figure*}

On the other hand, for less massive tracks ($< 2.5$ \Msun) 
there is always a coincidence of both significant ionisation rates 
and relatively large nebular masses, due to the 
slower evolutionary rate of the central star. However, the corresponding PNe
are still fainter in \oiiiline\ than the $2.5$ \Msun\ model because of
the lower ionisation rates.

In summary,  PN tracks with progenitors masses of $\sim2.5$ \Msun\ 
present the right combination of the dynamical timescale of the 
nebular expansion (that determines $M_{\rm sh}$ and $N_{\rm e}$), 
and  the evolutionary timescale of the 
central star (that determines $4\pi R^2Q({\rm He^+})$), 
to maximize the \oiiiline\ luminosity.

An additional effect in determining the PNLF cut-off 
is related to the PN lifetime. PNe produced by 
more massive progenitors stars are characterised by 
shorter $\Delta t^{\rm PN}$ and therefore, according to 
Eq.~(\ref{eq_massdist}), they contribute with fewer objects to the PNLF. 
In this regard, it should be noticed that, in their very early \Moiii\ 
evolution, some of the PN tracks of 
higher initial masses ($\ga4$~\Msun) 
may present spikes as bright as 
$\Moiii=-6.5$\footnote{These spikes show up 
mainly in the sets of PN models with $t_{\rm tr}=1000$~yr.}. 
However, these spikes do not affect at all
the position of the PNLF cut-off because they are 
too short-lived ($\la100$ yr).

Finally, we recall that 2.5~\Msun\ tracks correspond to  main sequence
lifetimes of just $7\times10^8$ yr. The minimum \Moiii\ 
magnitudes become progressively fainter as we go to tracks of
lower masses (larger turn-off ages). These facts have 
immediate implications for the dependence of $M^*$ on the 
galaxy's star formation history (Sect.~\ref{sec_ageburst}).

The next related question is now: What are the masses of the CSPNe 
forming the PNLF cut-off\,? 
According to our AGB calculations (Marigo et al. 1999) a 
star with initial mass of $\sim2.5$ \Msun\ and metallicity $Z=0.008$ 
ends its evolution as bare carbon-oxygen core of about $\sim 0.75$ \Msun\
(see Fig.~\ref{fig_ifmr}).
In Fig.~\ref{fig_mcsp} we show how the mass distributions of the CSPNe
are predicted to vary if a limit minimum \oiiiline\ magnitude $\widehat{M}$ 
is applied to the synthetic samples, i.e. we consider only PNe brighter than 
$\widehat{M}$. We see that at decreasing $\widehat{M}$, the peak of the 
theoretical distribution shifts towards larger masses, for less massive and 
fainter CSPNe fall outside the adopted luminosity cut.
As expected, in the extreme case of $\widehat{M}=-4.5$
(close to the observed PNLF cut-off) 
the peak is centered just at $M_{\rm CSPN}\sim 0.75$ \Msun 
(corresponding to the 2.5 \Msun\ progenitor star).
An interesting point to be noticed is that,
contrarily to what often assumed, the CSPN mass distribution
appears to differ from that of white dwafs (bottom-right
panel of Fig.~\ref{fig_ifmr}). The latter is in fact characterised by a 
sizeable tail towards larger masses, which is not expected in the case of
CSPNe. This can be understood by considering that the 
evolutionary speed of white dwarfs slows down so much that their lifetimes
can be considered practically infinite (regardless the mass of the stellar
progenitors), which determines a sort of accumulation of objects in the 
high-mass bins. 
Other differences between the mass distributions of CSPN and white dwarfs
may be introduced by observational biases, e.g. selection effects against
the detection of  PNe with slowly evolving, low-mass, central stars
(Stasi\'nska et al. 1997; see also Phillips 2000 for a critical analysis
of the empirical method used to evaluate $N(M_{\rm CSPN})$).

%%%%%%%%%%%%%%%%%%%%%%%%%%%%%%%%%%%%%%%%%%%%%%%%%%%%%%%%
\subsection{Dependence on H/He burning tracks}
\label{sec_HHetracks}

\begin{figure*}
\includegraphics[width=0.8\textwidth]{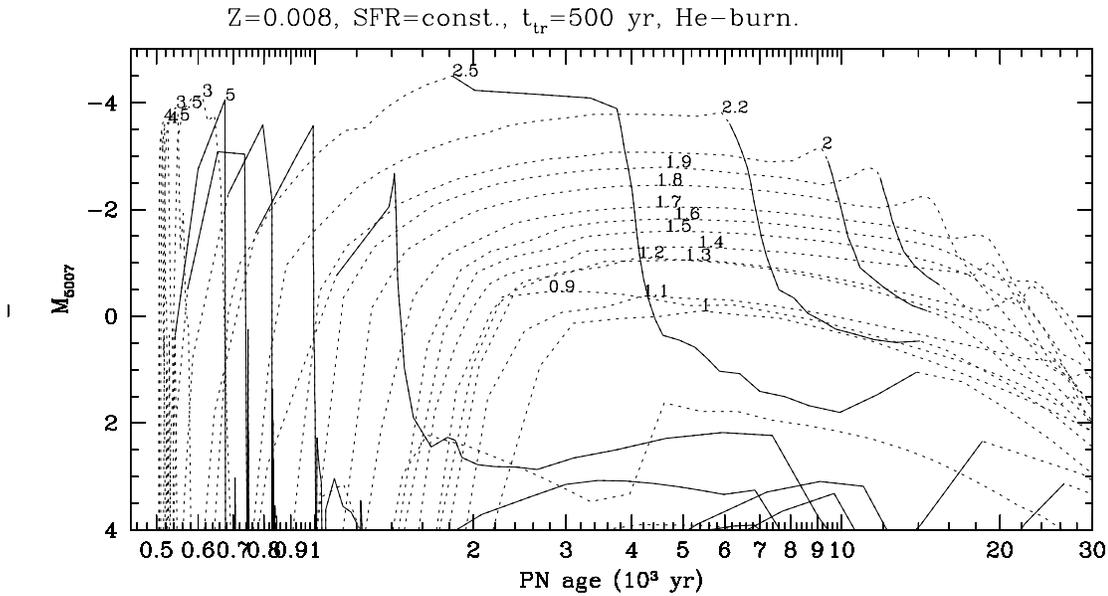}
\caption{The same as Fig.~\protect\ref{fig_whatcutoff}, but for
He-burning tracks. Notice also the wider range of PN ages compared to the
H-burning set}
\label{fig_whatcutoff_He}
\end{figure*}

As discussed in Sect.~\ref{sec_parameters}, the relative
proportions of PN H- or He-burning nuclei is essentially unknown.
An estimation is given by
Dopita et al. (1997) who derive a frequency ratio He/H $\sim 45\%-70\%$,
basing on evolutionary timescales and nebular sizes for a sample of 15 LMC PNe.

In general, 
He-burning tracks are characterized by longer evolutionary 
timescales and fainter luminosities than those of H-burning models. 
(see also Paper~I).
The resulting evolution of \Moiii\ is illustrated in 
Fig.~\ref{fig_whatcutoff_He}, that should be compared to the
corresponding figure for H-burning tracks of
Fig.~\ref{fig_whatcutoff}. It is evident that He-burning PNe
evolve slowlier, and preferentially in 
optically thin conditions.
However, the characteristic masses of the PNe at the cut-off 
change only slightly.

\begin{figure}
\includegraphics[width=\columnwidth]{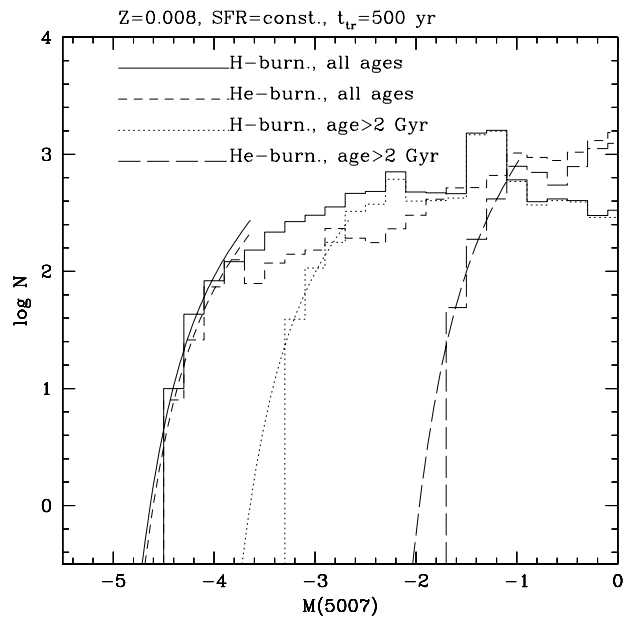}
\caption{Effect of the nuclear burning regime of
central stars (H- or He-burners) on the PNLF. {The histograms show
PNLFs as derived from simply counting PNe in 0.2-mag bins, whereas
the smooth lines show the result of fitting 
Eq.~(\protect\ref{eq_cutoff}) to the brightest 1-mag interval.}
For PN samples containing stars of all ages, the cut-off
is affected just slightly by the
nuclear burning regime of central stars 
-- in fact in the case here illustrated
$M^*$ is $-4.63$ for the H-burning case, and $-4.60$ for the He-burning 
one. Only when we limit to consider stars of older ages, the nature of the
tracks is of more relevance, He-burning PNe becoming much fainter
than H-burning ones: after 2 Gyr, $M^*$ becomes $-3.64$ and $-1.96$ 
for H- and He-burning tracks, respectively. }
\label{fig_PNLFhhe}
\end{figure}

Figure~\ref{fig_PNLFhhe} shows a direct comparison between 
PNLFs derived from H- and He-burning tracks, assuming  
$Z=0.008$, $t_{\rm tr}=500$~yr, and continuous SFR.
The shape of the PNLF somewhat changes, but not that dramatically, so that 
a fitting of  Eq.~(\ref{eq_cutoff}) to the brightest PNLF bins 
is straightforward. 
In fact, an unweighted least-squares fitting of 
Eq.~\ref{eq_cutoff}
-- performed on a PNLF with 0.2-mag wide bins, and 
limited to the brightest 1-mag interval -- 
produces $M^*=-4.63$ for the H-burning case, 
and $M^*=-4.60$ for the He-burning one, 
i.e.\ yielding almost identical values.
Similar results are obtained by using different transition
times and metallicities, while adopting either H- or He-burning tracks.

However, we note the $M^*$ tends to decrease much faster with age 
in the case of  He-burning tracks, than H-burning ones. This
effect can be appreciated in Fig.~\ref{fig_PNLFhhe}, and
is discussed further in Sect.~\ref{sec_ageburst} below. 

%%%%%%%%%%%%%%%%%%%%%%%%%%%%%%%%%%%%%%%%%%%%%%%%%%%%%%%%
\subsection{Dependence on the transition time}
\label{sec_ttr}

\begin{figure}
\includegraphics[width=\columnwidth]{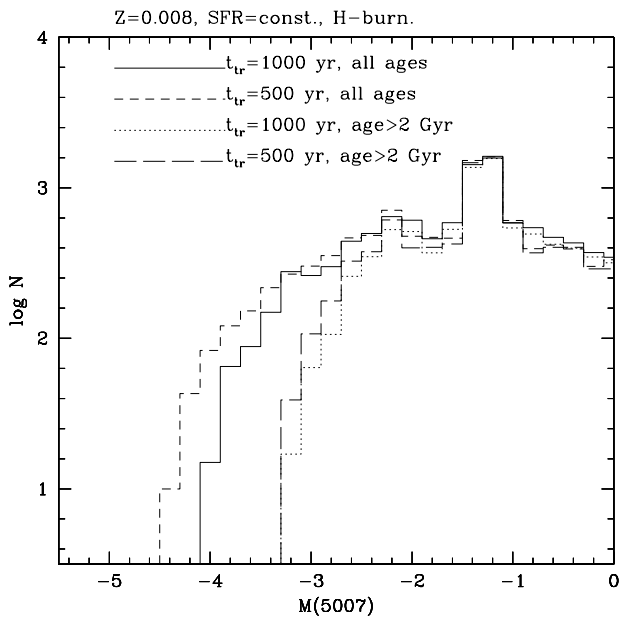}
\caption{Effect of the transition time on the PNLF, exemplified 
with the aid of the $Z=0.008$, H-burning models. 
For PN samples containing stars of all ages, the cut-off
is affected significantly by \ttr\ -- in fact $M^*$ changes from
$-4.63$ to $-4.33$ as \ttr\ varies from 500 to 1000 yr, 
in the first two cases here illustrated. When we limit only to stars of older 
ages, the effect of \ttr\ becomes of little importance.}
\label{fig_PNLFttr}
\end{figure}

Whereas there is no reason to believe that this is the case, 
the lack of robust predictions for the transition time 
(see Sect.~\ref{sec_parameters}) as well as  practical reasons   
impose that we first simulate PNLFs under the assumption \ttr\ 
is the same for all stars. In fact, when we do this
with our available tracks (for $\ttr=500$ and $1000$~yr),
we obtain the results depicted in Fig.~\ref{fig_PNLFttr}:
a high dependence of the cut-off magnitude $M^*$ on the transition time.
This can be explained by the large impact of \ttr\ on the
timescale for nebular expansion, especially for PNe of higher
masses, whose evolutionary timescales are comparable to \ttr\
itself. 

\begin{figure}
\includegraphics[width=\columnwidth]{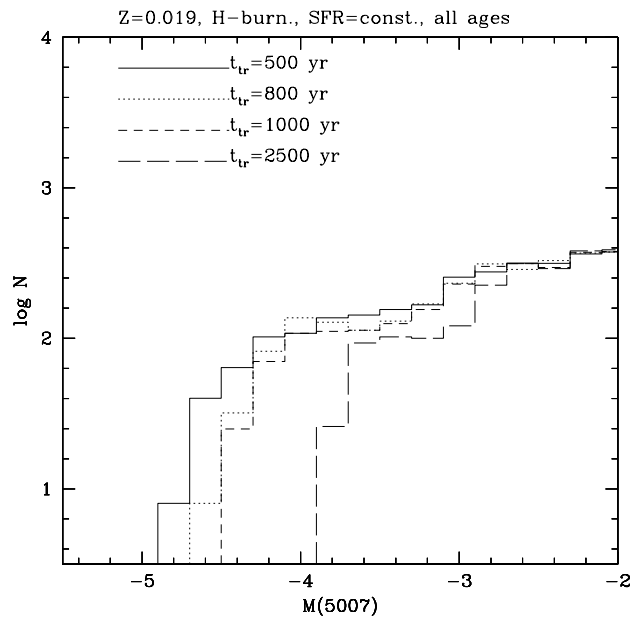}
\caption{The same as Fig. \protect\ref{fig_PNLFttr}, but
for $Z=0.019$ and 4 different values of \ttr, and limited to 
the case of continuous SFR. 
The trend of $M^*$ with \ttr\ is very regular, with values
$M^*=-4.81$, $-4.74$, $-4.67$, and $-4.23$ for $\ttr=500$, 800,
1000 and 2500 yr, respectively.}
\label{fig_PNLFttr2}
\end{figure}

In order to investigate further this behaviour, we computed additional 
sets of models for $Z=0.019$, H-burning tracks, and 
$\ttr=800$ and $2500$~yr (Fig.~\ref{fig_PNLFttr2}). 
If we consider PN samples that cover all ages, the cut-off
exhibits a very regular trend, becoming  brighter 
at shorter \ttr. The maximum \oiiiline\ brightness a PN sample may attain,
however, will be limited by the fact that \ttr\ cannot be shorter than a 
few hundreds of years. 
The value of $\ttr=500$~yr, succesfully tested in Paper~I and in this work 
on a number of Galactic PN observables (see Sect.~\ref{sec_comparison}),
is already an extremely low value if
compared to present results of evolutionary calculations (e.g. Vassiliadis
\& Wood 1994). 
Actually, it yields $M^*$ values 
that are slightly too bright, i.e. $\sim-4.8$ for  $\ttr=500$~yr, 
than the observationally-calibrated $M^* = -4.47$, 
while somewhat lower $M^*$ values are obtained with 
$\ttr=800$~yr and  $\ttr=1000$~yr.
On the other hand, \ttr\  longer  than 
$2500$~yr would produce too faint $M^*$ , i.e. $\ga-4.3$. 
In summary, according to  our present PN models, 
the observed cut-off magnitudes $M^*$,
for galaxies with ongoing SFR, will be obtained for \ttr\ values
between 500 and 1000 yr.

At this point two important remarks are worthy to be made.
First, as far as the bright PNLF cut-off is concerned, 
the indications we derive on  \ttr\ should be taken as a sort of calibration
that applies only to those PNe that actually form the cut-off, 
namely PNe with relatively
massive central nuclei ($M_{\rm CSPN} \sim 0.7-0.8 \Msun$), 
produced by relatively massive ($M \sim 2.5 \Msun$) and young 
($\tau^{\rm H}\sim 0.7$ Gyr) stellar progenitors.
In this range our results agree  with the predictions
for \ttr\  derived by Stanghellini \& Renzini (2000) from analytic relations
(see their figure 11).
It is clear that, more realistically,  $\ttr\ > 500-1\,000$ yr
should be associated to PNe with lower-mass stellar 
progenitors, but this would not affect at all our results 
on the cut-off brightness.

Second, a further support to our  
estimated ranges for \ttr\ and PN phase duration 
(see for instance Figs.~\ref{fig_whatcutoff} and \ref{fig_whatcutoff_He}) 
comes from 
the kinematic ages (from less of $1\,000$ yr up 
to almost $5\,000$ yr) empirically derived by Dopita et al.\ (1996)
for a HST sample of 15 PNe in the LMC. 
Many of these PNe are, in fact, classified
as Peimbert type I PNe, suggesting that they likely hide 
massive central stars. 

%Even if the range of \ttr\ values tested in this work is somewhat 
%limited, it already indicates that this is a critical parameter in
%determining the PNLF cut-off. 
%The observed cut-off magnitudes $M^*$,
%for galaxies with ongoing SFR, will be obtained for \ttr\ values
%between 500 and 1000 yr.

%%%%%%%%%%%%%%%%%%%%%%%%%%%%%%%%%%%%%%%%%%%%%%%%%%%%%%%%
\subsection{Dependence on age of the last burst of star formation}
\label{sec_ageburst}

\begin{figure}
\includegraphics[width=\columnwidth]{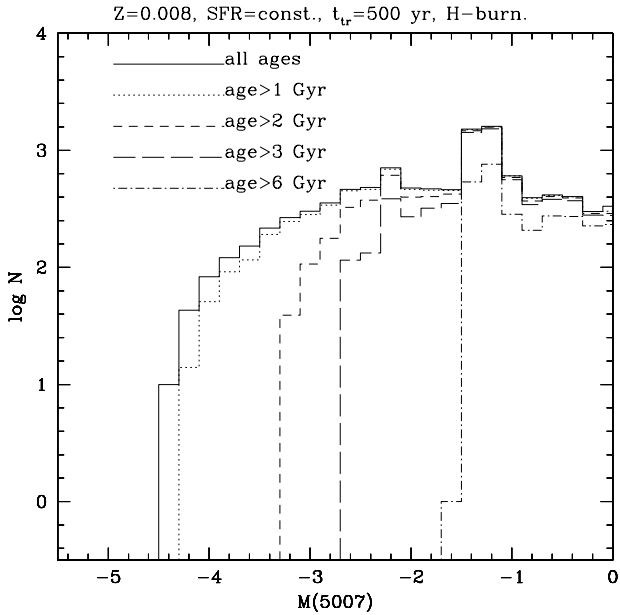}
\caption{Effect of the age distribution on the PNLF. To do this plot, 
we have generated a synthetic sample of PNe from a constant SFR, and then
considered in the PNLF only the PNe with ages older than the specified 
age limit. It may be seen that the cut-off
depends very much on the age of the latest star formation episode,
and hence on the recent (less than a few Gyr) star formation history.}
\label{fig_PNLFage}
\end{figure}

\begin{figure}
\includegraphics[width=\columnwidth]{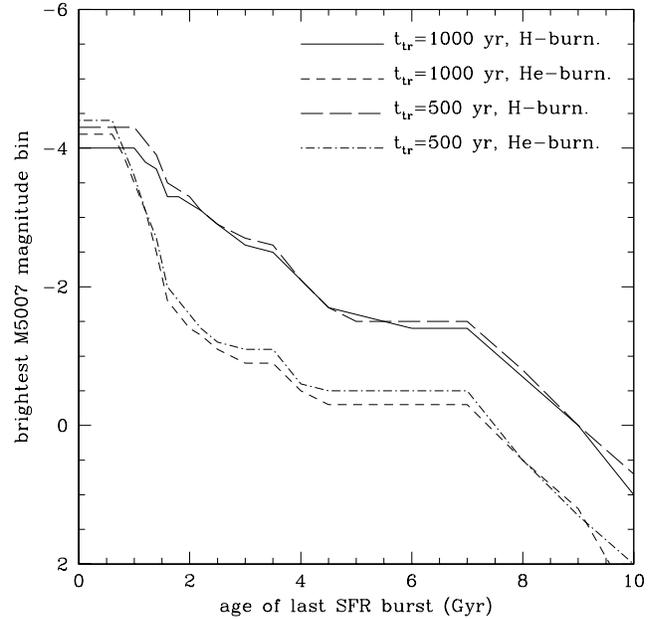}
\caption{Variation of the brightest magnitude bin
with the age of last SFR. This bin behaves similarly
to $M^*$ (although always $0.2-0.4$ mag fainter), 
but is more easily defined for all ages. 
The curve has been derived by searching for the
steepest change in the logarithmic of number counts,
in a series of PNLFs generated in the same way as in 
Fig. \protect\ref{fig_PNLFage}. Different cases (H-/He-burners,
two transition times) are shown, always for $Z=0.008$.}
\label{fig_cutoffage}
\end{figure}

The age of PN progenitors has long been suspected
to have a major effect on the cut-off magnitude 
(Dopita et al.\ 1992; M\'endez et al.\ 1993; Stanghellini 1995;
Jacoby 1996). 
Age is closely related to the turn-off mass, and to the
remannt mass via the IFMR.
For any reasonable relation of this kind, 
at older ages only PN central stars of lower mass, 
and hence of lower intrinsic luminosity, should form.
Our computations fully confirm the expected trends.

To illustrate the effect of the age, we simulate PN
samples adopting a SFR which is constant over the age range  
$[t_{\rm last},t_{\rm gal}]$. 
The upper limit, $t_{\rm gal}$, denotes the age of the oldest 
stars in a galaxy, and here it is assumed constant and equal to
12~Gyr. The lower limit, $t_{\rm last}$, is the age of the 
most recent burst
of star formation, and it is let vary between $0$ -- for
systems with ongoing star formation, namely the disk of
spirals and irregular galaxies -- and $t_{\rm gal}$ --
for predominantly old systems, like haloes and bulges,
and old ellipticals.
  
The results for the PNLFs are illustrated in 
Fig.~\ref{fig_PNLFage}. We clearly see that, as 
$t_{\rm last}$ increases, the PNLF becomes  progressively
depleted of its brightest PNe, and the cut-off 
shifts to fainter magnitudes. At ages older than a 
few Gyr, the PNLF may also change its the characteristic morphology,
i.e. shaping like a plateau with a
sharp cut-off in place of the smoother trend that is 
seen for younger $t_{\rm last}$. 

Figure~\ref{fig_cutoffage} shows the behaviour of the PNLF over the
brightest magnitude bin as a function of age. This bin
is simply defined as the one of the steepest change in the
logarithm of number counts. It behaves similarly to the 
cut-off $M^*$, but it is systematically fainter
than that quantity by 0.2 to 0.4 mag\footnote{The reason 
why we plot the brightest magnitude bin, is that it behaves
in a much smoother way than $M^*$, for all values of
$t_{\rm last}$. In fact, for $t_{\rm last}$ older than say 
4 Gyr, the PNLF often presents too sharp a cut-off 
followed by a shallow plateau, and a good fitting of 
Eq.~(\protect\ref{eq_cutoff}) is not possible.}.
As a matter of fact, the fading of the brightest magnitude
bin starts just after $t_{\rm last}$ becomes
older than the age of the brightest PN progenitors --
that means after $\sim0.7$~Gyr, which is the turn-off
age of $\sim2.5$~\Msun\ stars. Then, at older  $t_{\rm last}$, 
the dimming of the cut-off is clearly seen for all ages.

Finally, we remark that on the base of  our H-burning PN models, 
the fading rate of $M^*$ with $t_{\rm last}$ is of roughly $0.5$ mag/Gyr.
This rate is higher than those predicted by other
authors (e.g. Jacoby 1996) with the aid of simpler
models.

%%%%%%%%%%%%%%%%%%%%%%%%%%%%%%%%%%%%%%%%%%%%%%%%%%%%%%%%
\subsection{Dependence on metallicity}
\label{sec_metal}

\begin{figure}
\includegraphics[width=\columnwidth]{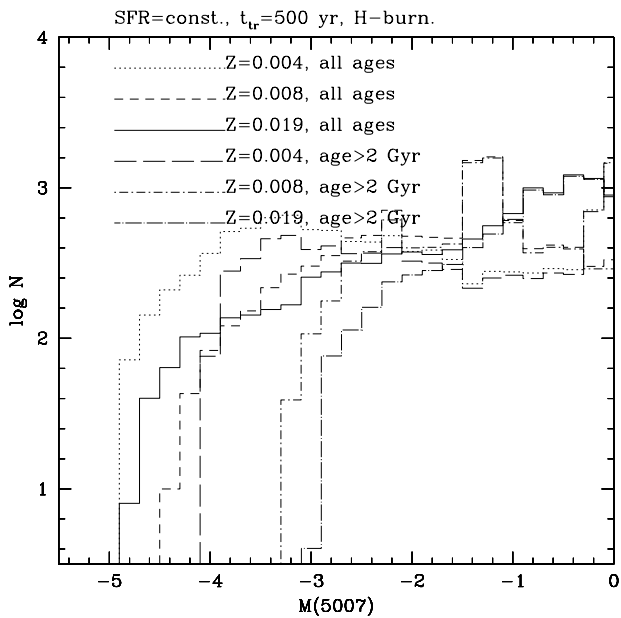}
\caption{Effect of mean metallicity on the PNLF.
For constinuous SFR, the cut-off depends on metallicity in a 
non-monotonic way. For older populations, instead, the behaviour
is monotonic. See text for details.}
\label{fig_PNLFmetal}
\end{figure}

Figure~\ref{fig_PNLFmetal} illustrates how the PNLF is predicted to
vary with metallicity according to our present PN models, 
for the cases of
$t_{\rm last}=0$ and $t_{\rm last}=2$~Gyr.
It can be noticed that our models exhibit an appreciable 
dependence of $M^*$ on metallicity. In the case of
continued SFR, this dependence is
not even monotonic with metallicity, being the $Z=0.008$
cut-off $\sim0.4$~mag fainter than those found for $Z=0.004$
and $0.019$. For $t_{\rm last}\ge2$~Gyr, instead, we find that
more metal-rich PN samples correspond to  fainter cut-offs.
The same behaviour is shared by all our sets of models.

\begin{figure}
\includegraphics[width=\columnwidth]{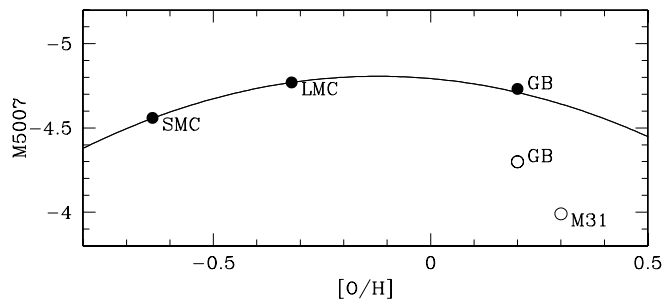}
\caption{Variation of $M^*$ with [O/H], according
to the quadratic relation (Eq.~\protect\ref{eq_OHciardullo}; 
continuous line) derived by 
Ciardullo et al.\ (2002) on the base of Dopita et al.'s (1992) data (circles).
Filled circles rely on the assumption 
that the stellar progenitors of the cut-pff PNe are 0.8~Gyr old, 
whereas empty circles show the location of the LMC and M31 if 
their SFR stopped 5~Gyr ago.
The adopted [O/H] values are taken from Dopita et al.\ (1992)}
\label{fig_dopita}
\end{figure}

These results may seem somewhat unexpected, considering 
the previous claims 
for a weak dependence of the cut-off on metallicity 
(e.g. Ciardullo et al.\ 2002). 
To understand this point,  first it is worth briefly recalling  
the origin of these claims. The most comprehensive analysis of the
\oh\ dependence is carried out by  Dopita et al.~(1992)
on the  base of photoionisation models of optically-thick nebulae.
It is found that the peak conversion efficiency of stellar photons 
to nebular \oiiiline\ ones is given by
	\beqa
\log\left\{\frac{F(5007)}{F({\rm star})}\right\}_{\rm peak} & = &
	-0.8791 + 0.1459\,\oh \\ \nonumber 
	& & - 0.3013\,\oh^2
	\label{eq_OHdopita}
	\eeqa
where \oh\ is the oxygen content in the nebula scaled to the Sun's.
By using this formula, together with  the inferred values for the 
absolute luminosity of the star at the PNLF cut-off
in a few galaxy systems (table 2 in Dopita et al.\ 1992), 
one can derive the following relation, 
quoted  by Ciardullo et al.\ (2002):
	\beq
\delta M^* = 0.928\,\oh^2 + 0.225\,\oh + 0.014
	\label{eq_OHciardullo}
	\eeq
which has indeed a weak dependence on the oxygen abundance,
as displayed in Fig.~\ref{fig_dopita} (filled circles). 
We should however emphasize that this latter relationship holds
only under a precise condition, namely the LMC, SMC, 
and Galactic Bulge contain young and massive PN progenitors, 
of ages similar to 0.8~Gyr.
If young stars were not present in the galaxies
studied by Dopita et al.~(1992), the fitting relation
by Ciardullo et al.~(2002) would not be valid, as it can easily be 
understood by looking at the empty circles in the figure.
Another interesting point is that, according to Eq.~(\ref{eq_OHciardullo}),
$M^*$ should reach a minimum for near-solar metallicities, 
contrarily to our findings.

Following Ciardullo (2003), 
the results of Fig.~\ref{fig_dopita} are the consequence of
two competing factors: (1) since oxygen is one of the main 
coolants of a nebula, the flux in \oiiiline\ increases only
with the square root of the oxygen content (Jacoby 1989); and
(2) the UV flux coming from the CSPN
decreases with metallicity. This latter trend, actually,
is dictated by the earlier onset of the superwind, and hence
the lower final stellar masses and luminosities, that
are expected in AGB models of higher metallicities.

After these considerations, how can we understand our 
results of Fig.~\ref{fig_PNLFmetal} in terms of 
basic stellar parameters\,?    
First of all, in agreement with other studies
(see e.g. Lattanzio 1986; Vassiliadis \& Wood 1993;
Wagenhuber \& Groenewegen 1998; Marigo 2001),
our underlying AGB models (Marigo et al.\ 1999)
present a clear trend of decreasing final masses at higher
metallicities, as depicted in  Fig.~\ref{fig_ifmr}. 
This behaviour ensures that, for given stellar age and initial
mass, lower-metallicity PNe will have central stars with  higher 
masses and hence UV fluxes, in agreement with  
Ciardullo's reasoning. 
We also remark that although the shape of the IFMR  
depends on a number of model details -- such as 
the presence of convective core overshooting in
previous evolutionary phases, the number and efficiency of
third drege-up events, etc. -- the key-role should be ascribed to 
the adopted prescription for mass loss 
as shown, for instance, by Groenewegen \& de Jong (1994).

The second important effect to be considered is the change of oxygen 
content \oh\ with iron content \feh. 
The initial main-sequence \oh\ values in our stellar models
are simply scaled to the Sun's determination\footnote{We recall that,
following  the usual designation, 
the abundance of a metal, $X_i$, is solar scaled 
when obeys the relation  $\textstyle{\frac{X_i}{Z}=(\frac{X_i}{Z})_{\odot}}$}
according to Anders \& Grevesse (1989), thus yielding 
$\oh=(-0.7,-0.4,0)$ for $Z=(0.004,0.008,0.019)$, respectively.
 
It is worth recalling that
stellar evolution models predict the surface oxygen abundance remains 
essentially unaltered after the first and second dredge-up episodes 
(see e.g. Girardi et al. 2000), while it may be affected to some extent
during the TP-AGB phase by the third dredge-up, and possibly 
by hot-bottom burning.

\begin{figure}
\includegraphics[width=\columnwidth]{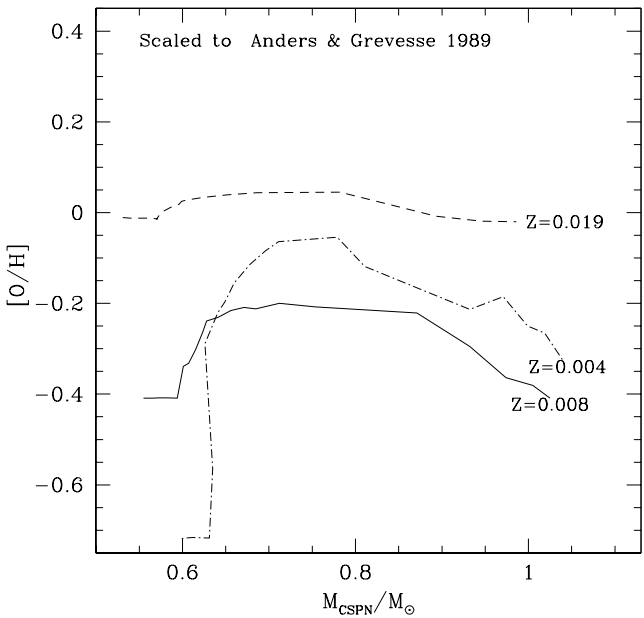}
\caption{Oxygen abundance at the end of the TP-AGB evolution
for the tracks used in this work, as a function of the core mass.}
\label{fig_oxygen}
\end{figure}

This latter process takes place in the most luminous and massive 
TP-AGB stars (with $M \ga 4 \Msun$) and it may deplete 
the stellar surface of oxygen (via the CNO cycle) 
in the case of high efficiency and sufficiently long duration 
of the CNO nuclear reactions.
This circumstance is not met in any of the TP-AGB models here considered.

On the contrary, the third dredge-up may cause the surface oxygen 
content to increase with time, together with the abundances of other elements 
like carbon and helium. The degree of oxygen enrichemnt in the envelope 
basically depends on three factors, namely: the efficiency of the dredge-up, 
the chemical composition of the dredged-up material, and 
the total number of dredge-up events experienced during the TP-AGB phase.  
Referring the reader to Marigo et al.\ (1999) and Marigo (2001) 
for more details about the TP-AGB models employed in this work,
we recall here that 1) the efficiency of the third dredge-up is parameterised
by a quantity $\lambda$, calibrated to reproduce the luminosity
functions of carbon stars in the LMC and SMC; 
the oxygen abundance in the dredged-up material
is assumed of just $X(^{16}{\rm O})\sim 0.02$ (in mass fraction), 
that represents 
the standard order of magnitude obtained by most nucleosynthesis 
calculations of thermal pulses (e.g. Boothroyd \& Sackmann 1988; 
Karakas et al. 2002; 
but see also Herwig et al. 1997 for different results).
  
Despite this rather low value, it turns out (see Fig.~\ref{fig_oxygen}) that
the predicted \oh\ abundances at the end of the TP-AGB 
-- and during most of the PN phase -- may  significantly differ from
the initial main-sequence values, the deviation being more pronounced in 
models of lower metallicities because of the higher dredge-up efficiency 
and larger number of thermal pulses.
Moreover, we notice that \oh\ does not vary at all at 
lower initial masses, say $M\la1.2-1.5$~\Msun (depending on metallicity),
since the third dredge-up is not predicted to take place in those models.

\begin{figure}
\includegraphics[width=\columnwidth]{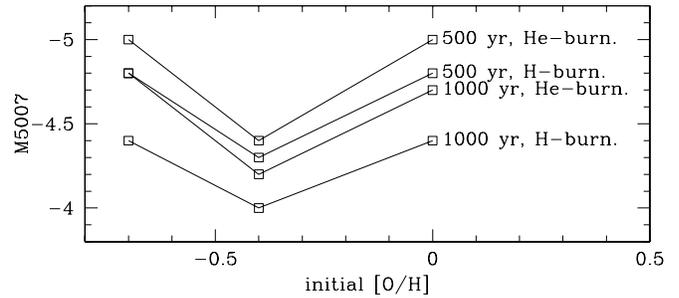}
\caption{Expected variation of  $M^*$ with the initial main-sequence [O/H]
values, according to our sets of models 
(with parameters $\ttr=500$ and 1000~yr, and H-/He-burning CSPNe), 
as indicated (empty squares). 
We recall that the [O/H] ratio predicted in our PN models may
be quite different from the assumed main-sequence values. 
See Fig.~\protect\ref{fig_oxygen} and text for more details}
\label{fig_ours}
\end{figure}

Figure~\ref{fig_oxygen} clearly shows that the assumption of
$\oh=\feh$, or that \oh\ is the same as in the \hii\ regions of a galaxy,  
may be not suitable for PNe -- unless oxygen dredge-up does not occur
in AGB stars, which is unlikely considering our present understanding 
of stellar evolution. Moreover, the change of \oh\ 
is not necessarily monotonic with \feh; in fact most of our models
with $Z=0.004$ present \oh\ values higher than those
found at $Z=0.008$, but still lower than those for $Z=0.019$.
A monotonic behaviour of \oh\ with $Z$ (and equivalently with
\feh) is recovered only by models with lower masses -- hence older progenitors'
ages --, that do not experience the third dredge-up.

\begin{figure}
\includegraphics[width=\columnwidth]{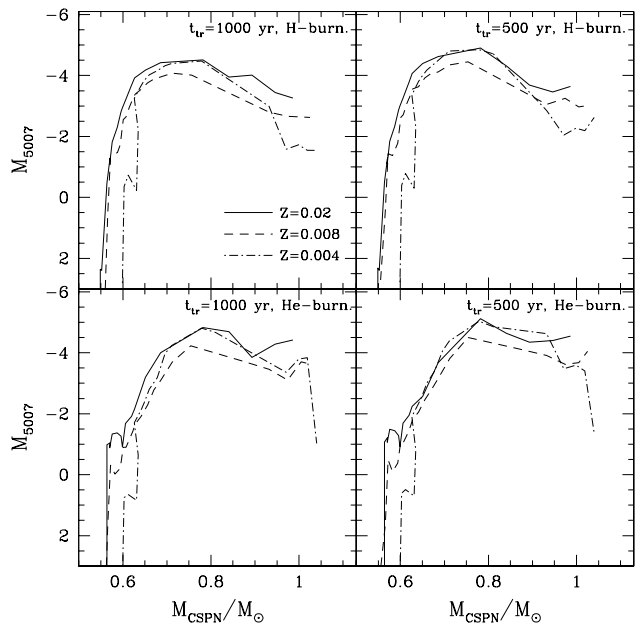}
\caption{Peak \Moiii\ magnitude for our models, for the several
cases explored and as a function of CSPN mass. Notice
the similarity with the trend of \oh\ with metallicity, shown in
the previous Fig.~\protect\ref{fig_oxygen}.}
\label{fig_M5007_peak}
\end{figure}

In summary, 
if one accounts for the changes of  \oh\ at the surface 
of  stars with different masses and metallicities in the framework 
of the present stellar evolution models, the resulting  
dependence of \oh\ (in PNe) with metallicity cannot be expressed by 
a simple scaling relation, rather it is characterised by a more complex 
behaviour as, for instance, illustrated in Fig.~\ref{fig_oxygen}.

Such behaviour gives origin to the non-monotonic trend of 
$M^*$ with \oh\ displayed in Fig.~\ref{fig_ours} (empty squares) which,
in turn, reflects in the non-monotonic trend of 
$M^*$ with age. In order to illustrate it better, 
Fig.~\ref{fig_M5007_peak} shows the peak \Moiii\ magnitude
for all our models, as a function of CSPN mass.
The similarity of the curves with the behaviour of \oh\
shown in Fig.~\ref{fig_oxygen} is evident.

To further investigate the sensitivity of $M^*$ to the actual O/H in PNe,
we have performed a test on our set of $Z=0.019$, $\ttr=500$ yr,
H-burning models, simulating a lower main-sequence oxygen abundance.
In practice, we have re-calculated the photoionisation structure  
of these PN models providing CLOUDY with an input \oh\ lowered by a 
constant quantity of $-0.24$ dex\footnote{
The choice of this value is motivated by the recent re-determination 
of the Sun's oxygen 
abundance by Allende Prieto et al.\ (2001) that is lower by just $-0.24$ dex 
than previously estimated by Anders \& Grevesse (1989).}
, so that the \oh\ curve for $Z=0.019$
in Fig.~\ref{fig_oxygen} would be simply shifted downwards by this amount 
while keeping its shape unaltered. 
It follows that the new  \oh\  curve for $Z=0.019$ would come closer 
to that for the $Z=0.008$ set at the typical CSPN masses
($\sim 0.70-0.75 \Msun$) of the cut-off PNe
(in the case of continous SFR, or equivalently for $t_{\rm last}=0$). 

With the aid of PNLF simulations for the new oxygen abundances 
we obtain that such decrease in \oh\ causes
a dimming of the PNLF cut-off amounting to $0.48$ mag 
for $t_{\rm last}=0$, and to $0.43$ mag 
for $t_{\rm last}=2$ Gyr. These numbers correspond to 
an incremental ratio $\Delta M^* / \Delta \oh$ of $\approx 
2$ mag dex$^{-1}$, which predicts the right
order of magnitude required to explain our results of
Figs.~\ref{fig_PNLFmetal}, \ref{fig_dopita}, 
and \ref{fig_M5007_peak}, i.e. in particular the difference in $\Delta M^*$
between the PNLFs calculated at different metallicities.

This significant sensitivity of \Moiii\ to \oh\ displayed by our 
models raises a few considerations of general validity. 
In fact, it appears clear that the predicted values of \Moiii\  
should crucially depend on:
\begin{itemize}
\item
 The assumed value for the initial main-sequence oxygen content,
hence the adopted solar oxygen abundance (if solar-scaled compositions
are assumed, as usually is the case).
We remark that the determination of the Sun's oxygen abundance 
 has changed dramatically over the years --
i.e. from $\log({\rm O/H})=-3.07$ in Anders \& Grevesse (1989) to
$\log({\rm O/H})=-3.31$ in Allende Prieto et al.\ (2001). 
This alone would cause a change of $0.24$ dex  
in our scale of oxygen abundances.
 As mentioned above, a systematic decrease of $-0.24$ dex in our 
O abundances -- necessary if we were to bring our models to
the solar-composition scale of Allende Prieto et al.\ (2001) 
-- would cause a dimming of $\sim+0.4$ mag in our $M^*$ values.
\item 
(2) The choice of the input TP-AGB models in the PNLF simulations.
In particular, the predictions of  \oh\ in PNe are  importantly
affected by several (often poorly known or improperly assumed) model details
such as: inter-shell nucleosynthesis during thermal pulses,
efficiency of the mixing episodes, mass-loss prescrition,
assumed molecular opacities, etc. 
\end{itemize}

Finally, 
we recall that there is an additional reason that likely concurs 
to produce the differences in the predicted metallicity dependence 
of the PNLF cut-off observed between ours and the 
one derived from Ciardullo et al.\ (2002) on the base of
Dopita et al.'s (1992) work. In fact, while we find that 
both optically-thin and thick PNe should populate the
PNLF cut-off, the analysis by Dopita et al. considers only the
behaviour of optically thick objects.

%%%%%%%%%%%%%%%%%%%%%%%%%%%%%%%%%%%%%%%%%%%%%%%%%%%%%%%%
\section{Discussion}
\label{sec_discussion}

\subsection{Summary of main dependences}

In this paper we have carried out a theoretical investigation of the 
\oiiiline\ PNLF and its sensitivity to various stellar evolution and 
population parameters. To this aim a detailed photoionisation code 
(CLOUDY) has been incorporated into our synthetic PN model (Paper~I),  
so that we can rely on a simple but quite complete picture of PN evolution, 
which allows a satisfactory reproduction of many PN observables 
like well-known correlations involving nebular radius, ionised mass, 
density, intensity line ratios, etc.
 
In particular, major attention has been paid to the modelling of the bright
cut-off of the PNLF, that is claimed to be a powerful distance 
indicator in virtue of its observed invariance in host galaxy systems of 
different ages and metallicities.
The results of our analysis on the PNLF cut-off 
can be summarised as follows. In order of decreasing importance, the 
main parameters affecting the cut-off magnitude $M^*$ are: 

\paragraph{The age of the last burst of star formation, $t_{\rm last}$.} 
It turns out that, over the entire range of old and intermediate ages
(say from 13 to 0.1 Gyr; see Fig.~\ref{fig_PNLFage}), 
the possible brightest PNLF cut-off is obtained and  
remains invariant as long as  $t_{\rm last} \la 0.7$ Gyr, that is if 
stars with initial masses $\ga 2.5 \Msun$ (and final masses $\ga 0.75 \Msun$)
are presently evolving through the PN phase.
At older ages, for $t_{\rm last} \ga 1$ Gyr, we predict a sharp dimming of 
the PNLF cut-off, as it becomes populated by much fainter PNe 
evolved from lower mass progenitors. 
%A second, rather extended, 
%but very faint plateau in the $M^*$ vs. $t_{\rm last}$
%diagram is reached at ages roughly from  4 to 7 Gyr.
%For the cases here considered, the invariance of the cut-off 
%over this age range can be explained by considering that the progenitor 
%stars are all expected to form PN nuclei of approximately the same mass,
%at around 0.6 \Msun. ****
 
\paragraph{The transition time, $t_{\rm tr}$.} 
This is one of the most ill-known quantities on the theoretical ground, 
mainly reflecting major uncertainties about stellar mass loss during 
the initial post-AGB stages. In addition, $t_{\rm tr}$ should 
be affected by large variations depending on stellar mass and phase 
of the last thermal pulse on the AGB.
Given these premises, our assumption of constant transition time may appear 
too simplicistic, nevertheless we are still able to derive a few 
relevant indications, namely:  
i) In the case of continuous SFR, the PNLF cut-off depends only on 
$t_{\rm tr}$ of those CSPNe effectively populating it 
(with masses around $0.7-0.75$ \Msun). A satisfactory reproduction 
of the observed cut-off brightness is obtained with
$t_{\rm tr} \sim 500-1000$ yr.
Conversely, the details of the PNLF cut-off are insensitive
to $t_{\rm tr}$ of fainter, slowly evolving PNe
originated from lower-mass progenitors. 
ii) In the case of $t_{\rm last} \ga 1$ Gyr the dependence of the
cut-off on $t_{\rm tr}$ is weak. In fact, this time 
the cut-off PNe correspond to lower-mass central stars, 
for which the condition $\Delta t^{\rm PN} \gg t_{\rm tr}$
is usually met for reasonable values of the transition time.
In other words, the effect of  $t_{\rm tr}$ on the PNLF cut-off is
practically smeared out by the long evolutionary timescales  of the PN phase.

\paragraph{The nuclear burning regime of the central star.} 
Unfortunately, the relative contributions of H- or He-burning central stars 
to a PN population are still poorly determined.
With the aid of PNLF simulations obtained assuming that all CSPNe are either   
H- or He-burning central stars, we find that variations in the 
the cut-off brightness remain moderate at 
younger ages, instead becoming huge at older ages.
In general, we find that the PNLF cut-off for H-burning central stars 
is brighter than that for the He-burners, since these latter are  
characterised by slower evolutionary rates at fainter luminosities.  
Anyway, it is reasonable to think that both H- and He-burning 
central stars will be present in the observed galaxies, so that
the brightest ones -- i.e. the H-burning -- may actually 
determine the observed PNLF cut-offs, as well as the age-sensitivity
of the sample. 

\paragraph{The metallicity.} 
Instead of metallicity, we should refer, more correctly, to 
the dependence on the oxygen content \oh\ . Oxygen  actually plays a major
role in the nebular cooling, and its abundance directly affects 
the intensity of the \oiiiline\ line.
The oxygen content in PNe traces 
the metallicity of the parent galaxy only assuming that the  
PN stellar progenitors keep their initial surface oxygen abundance unaltered
during the entire evolution up to the PN ejection.
In the framework of present stellar evolution models, some increase of
oxygen may be produced by the third dredge-up occuring during the TP-AGB
phase, the degree of enrichment being more pronounced at lower metallicities.
Moreover, the predicted \oh\ in PNe depends on the assumed initial composition
of the gas, that is usually assumed  to be solar-scaled ($\oh=\feh$).
All these factors yields a metallicity dependence of the cut-off that 
is not monotonic and also changes with age.

\subsection{On the claimed age invariance of the PNLF cut-off}
 
After this brief recapitulation of the main dependences in the models, 
let us now discuss to more detail the claimed 
age-invariance of the cut-off.
For the cases with continued SFR, our models predict $M^*$ values
that are very close to the observed one for the LMC
and other similar galaxies in the Local Group. 
This, and the strong decay of $M^*$ with age,
would suggest that we have a natural explanation for
the cut-off position in star-forming galaxies, 
it being determined
by stars with initial mass of $\sim2.5$~\Msun. This is
also the conclusion reached by Dopita et al.\ (1992),
based on a detailed comparison between PN models and
observations in the Magellanic Clouds.

However, considering that sizeable changes of $M^*$ are expected 
depending on various model details, and mostly on the age of PN progenitors,
how can we reconcile such predicted  variations with the observational 
indications of a fairly constant PNLF cut-off brightness, common to  
galaxy systems of all types\,?
In fact, an important point is that  
elliptical galaxies in groups appear to share similar cut-offs
with spiral galaxies.

At this stage, it is worth  recalling and discussing the 
different interpretations
to this issue that have been proposed in the recent literature, referring
mostly to Jacoby (1996) and Ciardullo (2003), who provide two extended 
reviews on the PNLF as a distance indicator. Actually, the two reviews 
contain two different, perhaps complementary,  
explanations to the age-invariance of the cut-off brightness.

Following Jacoby (1996, and references therein) the constancy 
of the PNLF cut-off, 
over the age range of $3-10$ Gyr, should be essentially ascribed 
to the extremely narrow 
distribution of central star masses, say $0.58 \pm 0.02$ \Msun. 
This would correspond to
initial masses in the range $1-2$ \Msun\  according to the semi-empirical 
IFMR for the solar neighbourhood (Weidemann 1987; see also Herwig 1997).  
Our AGB models for solar metallicity  (see Fig.~\ref{fig_ifmr})
recover well this part of the observed IFMR, 
nevertheless the present PNLF simulations predict, on the contrary, 
large variations of the cut-off over the above age range.
Actually, the cut-off brightness seems somehow to flatten over 
the age range $3-7$ Gyr 
(see  Fig.~\ref{fig_PNLFage}), but such plateau is incurably 
fainter than the observed
cut-off. Moreover, we notice that metallicity effects are likely to 
affect the IFMR, i.e. larger final masses at lower metallicites. 

Ciardullo (2003), on the base of Ciardullo \& Jacoby's (1999) paper,  
suggests that the stringet correlation between 
mass of the central star and  circumstellar extinction 
(due to dusty envelopes ejected
during the AGB phase) should prevent high-core mass PNe from 
populating brighter bins
beyond the observed PNLF cut-off. In other words, over-luminous 
Population I PNe (evolved from young and massive progenitors, 
hence potentially detectable as very  
bright \oiiiline\ sources in spiral and irregular galaxies) 
would be efficiently extinguished below the observed cut-off.

To this respect we notice that there is an additional factor, 
other than extinction,
that would locate high-core mass PNe at fainter magnitudes than the cut-off.
To this respect, we should consider that
one of the underlying assumptions in Ciardullo \& Jacoby (1999) analysis 
is that a constant fraction of the stellar luminosity
is reprocessed into \oiiiline\ emission, which yields 
a positive correlation between
the mass of the central star and  the maximum \oiiiline\ luminosity emitted
during the PN phase.
Our calculations, instead, show that this correlation is not monotonic, 
as the maximum \oiiiline\ luminosity presents a peak at central star 
masses of about $0.70-0.75$ \Msun, then declining at larger masses.
This implies that, even without the effect of circumstellar extinction, 
PNe with high-mass central stars would not contribute to the PNLF 
cut-off because intrinsically fainter.

In conclusion, we should stress that our difficulty to explain 
the age-invariance of the PNLF cut-off could not be solved
by including the effect of circumstellar extinction, as proposed
by Ciardullo \& Jacoby  (1999). 
In fact, we do not have any need to
make PN models with massive progenitors fainter, as required by
Ciardullo \& Jacoby (1999), but rather to make PN models
with less massive and old  progenitors several magnitudes brighter. 
How to obtain this effect with the aid of consistent PN models, 
is still to be investigated.

\begin{figure}
\includegraphics[width=\columnwidth]{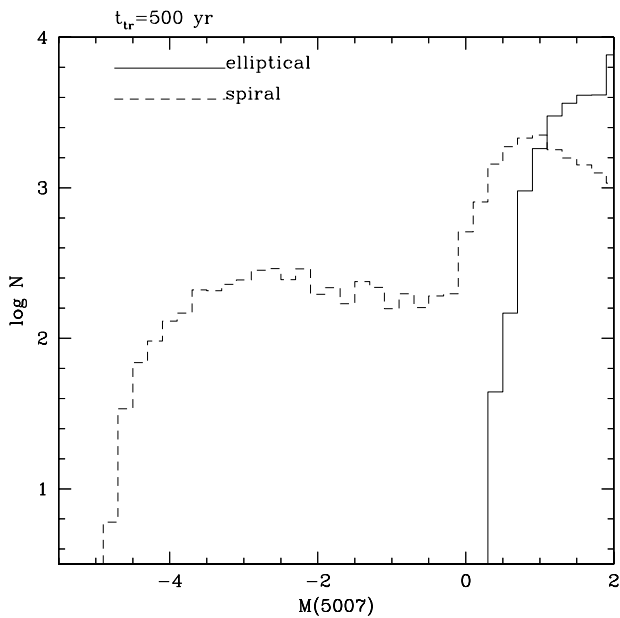}
\caption{The PNLFs for 2 template galaxies: a massive elliptical and
a spiral one. See the text for more details.}
\label{fig_PNLFtemplate}
\end{figure}

\subsection{The dependence on galaxy type: the age alternative }

The PNLF is expected to depend on the galaxy type and even more 
relevant here on the age of the stellar populations.
In order to discuss this point, we start by presenting in
Fig. \ref{fig_PNLFtemplate} the predicted
PNLFs for two template galaxy models representing two 
extreme cases: massive ellipticals and spirals. 
The elliptical galaxy corresponds to the $10^{12}$ \Msun\ model 
from Chiosi \& Carraro (2002), in which the star formation is all 
concentrated in the first 2 to 4 Gyr of the galaxy life; during
the same period its metal content rapidly increases up to 
near-solar values.
The spiral galaxy model, instead, is derived from chemical 
evolution models of our own Galaxy: it corresponds to model B from
Portinari et al. (1998). The star formation in this case is 
characterized by a initial rapid increase during the first 3 Gyr,
followed by ever-decreasing, but continuous, SFR. The AMR 
is derived from models but is in excelent agreement with 
observational data from disk stars.

These template galaxy models show clearly a difference of 
about 5 mag between the cut-off position expected (and observed)
in spirals at $M^*\simeq-4.5$, and the one expected in old ellipticals
at $M^*\simeq0.5$. This difference follows, essentially, from
significant age dependence expected by our models 
(see Fig.~\ref{fig_cutoffage}), and is in clear contrast with
observational claims (see Sect. \ref{sec_intro}).

\begin{figure}
\includegraphics[width=\columnwidth]{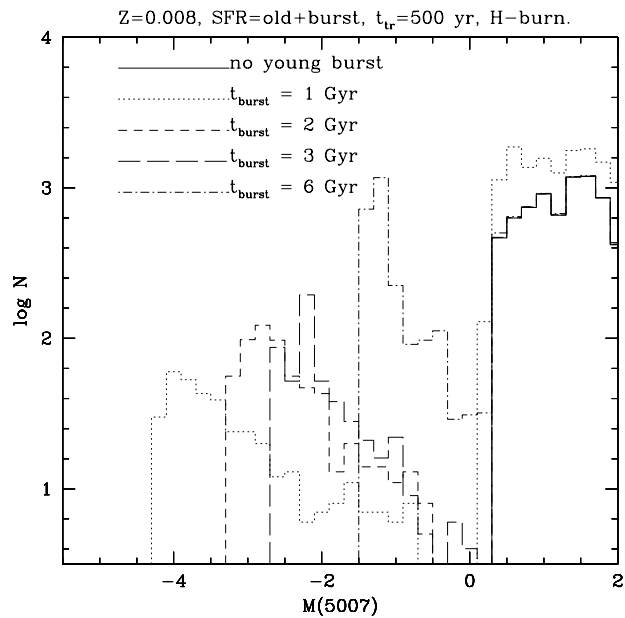}
\caption{The same as Fig.~\protect\ref{fig_PNLFage},
but now simulating SFR with just 2 bursts of SFR: an old an intense 
burst in the beginning of the galaxy history, of age 10 Gyr, 
added to a more recent burst just 5 percent as intense as the old one.
It can be seen that the PNLF shapes change a lot, with respect to
the continuous SFR shown in Fig.~\protect\ref{fig_PNLFage}. 
However, the bright PNLF cutoff is found in the
same magnitude bins as before.}
\label{fig_PNLFburst}
\end{figure}

Thus, if ellipticals were systems consisting exclusively of old stellar 
populations, we could hardly explain the invariance of the cut-off.
A viable explanation for the observed invariance   
is that the ellipticals with measured PNLFs may have had a 
burst of star formation in recent times, with ages close to 
1 Gyr. In order to illustrate this possibility, 
in Fig.~\ref{fig_PNLFburst} we present a series of simulated
PNLFs for galaxies having an intense burst of SFR
10 Gyr ago, followed by a more recent one at the age $t_{\rm burst}$,
which involved just 5 $\%$ of the stellar mass in the galaxy.
It can be noticed that the shape of these PNLFs does change a lot,
compared to the continuous-SFR PNLFs illustrated in 
Fig.~\ref{fig_PNLFage}. Actually, in the cases of bursting SFR, 
the PNLFs present a brighter component, due to the more 
recent burst, that  displays a cut-off behaviour over the brightest bins,
while  continuously declining towards fainter magnitudes.
Then, at about $\Moiii\ga0.3$, a second, highly-populated and faint
component builds up as a consequence of the old burst.
We also notice that if just 
the brightest PNe were observed in such galaxies, the derived
cut-offs would be exactly those caused by the youngest
burst. Only deeper observations could then reveal the 
different shape of these brighter components, and the sharp rise -- 
actually a second cut-off -- at $\Moiii\ga0.3$.
Therefore, if PNLFs are observed deeply enough they can give 
very valuable results about the star formation history during 
the last several billion years.

In summary, in the context of our PN models, the formation of  
the PNLF cut-off, observed in some elliptical galaxies as bright   
as in spirals, could be explained only invoking a recent burst of SFR, 
as young as $\approx 1$ Gyr, corresponding to stellar progenitors 
with initial mass 
of about $2.5-3.0$ \Msun, or correspondingly to CSPN mass of about 
$0.70-0.75$ \Msun.

Let us also clarify that a change of ill-known model parameters 
hardly would change this scenario. In fact, in order to reconcile our 
models with the claimed age-invariance of the PNLF cut-off, we
would have to get much brighter PNe models at old ages. By using He-burning
post-AGB tracks instead of H-burning ones, our old models become
even fainter than above mentioned. Also, older PN models are 
little sensitive to the exact value
of the transition time, and choosing these times much larger than
500 yr -- as likely appropriate for models of lower mass -- 
would also cause fainter old PNe models, instead of brighter ones. 

The idea the early type galaxies may suffer secondary recent  
star formation activity has long been around in literature. It is 
hinted by the diagnostics of
the line absortption indices on the Lick system, such as $H\beta$ and/or 
$H\gamma$, that are commonly considered as good 
indicators of the age of the most recent star forming episode
(Bressan et al. 1996; Jorgensen 1999; Kuntschner 2000;
Longhetti et al. 2000; Trager et al. 2000a,b; Vazdekis et al. 2001; 
Davies et al. 2001; Poggianti et al. 2001; Tantalo \& Chiosi 2003),
Further supported by evidences from various broad-band colors 
(Zabludoff 1996; Goto et al. 2003; Quintero et al. 2004; 
Fukugita et al. 2004), and finally found also 
in NB-THSP simulations of galaxy formation and evolution 
(Kawata 2001a,b; Carraro \& Chiosi 2003; 
Kawata \& Gibson 2003a,b). However, the 
constraint set by the PNLF cut-off luminosity is very 
demanding because it would imply that
in all ellipticals showing bright PNs star formation activity
should have occurred as recently as about 1 Gyr ago in order to reach 
the same luminosity of the ongoing star formation case as in spirals. 

The question may be posed whether sampling ellipticals and 
spirals to derive the empirical PNLF, 
selection effects somehow favour those ellipticals with 
bright PNe, i.e. recent star formation, so that ellipticals 
with older stellar content and whence fainter cut-off luminosity 
for their PNLF are simply missed.
This issue is not specifically mentioned in literature so that 
a clear answer cannot be given here. 

The most puzzling aspect with the age interpretation is that there 
are some elliptical galaxies,
for instance  NGC4649, NGC4406, NGC4486, NGC1399, and NGC1404, 
which have bright PNs but yet 
possess broad colors and line absorption indices -- 
for instance mean $B-V = 0.96$ and mean $H\beta= 1.45$ -- strongly 
indicating that only old stars are present, unless the last episode has  
involved less 0.05\% of the galaxy mass as thoroughly discussed by 
Tantalo \& Chiosi (2003).

{A possible way to remove the theoretical discrepancy might be 
related to the IFMR and its dependence on metallicity. In practice, 
we could obtain much brighter PNe at old ages if their low-mass and 
metal-poor stellar progenitors had been able to build up relatively
massive cores at the end of the AGB evolution, hence brighter CSPNe.
Indeed a general trend of increasing final mass at decreasing $Z$ 
is already predicted by our models (see Fig. \ref{fig_ifmr}), but the
resulting effect on the PNLF cut-off is not strong as the 
lowest metallicity here considered is just $Z=0.004$.
Additional calculations at lower metallicity are then required to 
test the above hypothesis. This analysis is in progress.
Alternatively, }
we are left with the open possibilities that either the current 
models of PNs are still too 
simple and neglect some important effect (we are, however, 
confident that this is not the case), or 
some new type of old stars intervene in particular in the 
case of old elliptical galaxies in such 
a way they closely mimic the properties of classical PNs 
(investigation of this topic is currently underway).

\section{Concluding remarks}
\label{sec_conclusion}

In Paper~I we had developed a theoretical model for synthetic PNe. 
In the present paper we use this model to predict properties of 
synthetic samples of PNe. In particular we investigated the PNLF 
of synthetic PN populations with regard to its shape, 
maximum \oiiiline\ brightness, 
and how it depends on age and composition of the population as well 
as on assumptions necessarily entering our model. 
With respect to Paper~I and in view of the importance of a 
reliable prediction of the 
emission properties of PNe, we extended our model by including 
%the radiation transfer 
the photoionisation 
program CLOUDY, thereby eliminating one of the free parameters, 
the electron temperature 
\Te\ in the nebulae. The detailed radiation transfer treatment 
confirms that the assumption 
of an approximately constant \Te\ of about $10000\,\rm{K}$ in the
dynamical part of the model is justified, at least 
within the accuracy required at this stage. Apart from free 
assumptions about age 
and composition of the population causing PNe, and about the 
SFH and IMF, the transition 
time remains the sole physical parameter for which we have 
only rough indications, and 
for which we would need additional theoretical calculations 
of the superwind phase of AGB stars,
and their transition to the post-AGB phase. We are also forced 
to assume spherical symmetry of the circumstellar matter.

With this improved model for PNe populations we have simulated 
galactic PNe and 
compared the results with observations. We concentrated on 
emission lines and ionisation states, 
and find excellent agreement without the necessity to vary 
any model parameters. Based on these 
successful comparisons, we then concentrated on the 
luminosity function. The main results concerning the PNLF are:
\begin{itemize}

\item Our models predict 
a PNLF shape towards the cut-off very similar to those observed
in spiral and irregular galaxies. 

\item In galaxies with recent star formation (either continued up to now,
or with a burst occurred $\sim 1$ Gyr ago), the cut-off
is expected at $-5 \le \Moiii \le -4$, depending 
on model parameters. This is consistent with the observed values at
about $-4.5$ (e.g. for the LMC).

\item In galaxies without recent star formation, 
the cut-off is predicted much fainter than observed.
This is due to the lack of stars with $M>2$ \Msun, 
responsible for the brightest PNe ($\Moiii\le-3.0$).

\item  From the two previous points it follows that we can hardly 
explain the constancy of the cut-off observed in some ellipticals (dominated
by old populations) in terms of the claimed age-invariance. 
Rather, a possible way to explain the bright  
PNLF cut-off observed in some late-type galaxies
is just to invoke a recent
burst of star formation.

\item One  main parameter determining the PNLF cut-off is  
the transition time, to be  
constrained on the base of observed PN properties 
(e.g. $N_{\rm e}$, $T_{\rm e}$, ionized masses, radii, 
expansion velocities, etc). 
Our simulations indicate that, in order to recover the observed brightness
of the cut-off,  the  CSPNe populating it (with masses around 
$0.70-0.75$ \Msun) should have a \ttr\ of the order of few
hundred years (typically $500-1000$ yr). On galaxies without recent SFR,
the cut-off position becomes little sensitive to \ttr.

\item The predicted dependence of the PNLF on metallicity is the result 
of a number of factors, as we need to know (i) 
how the evolutionary properties of the stellar
progenitors (i.e. the efficiency of mass loss on the AGB, the details
of the initial--final mass relation) vary as a function of the initial
chemical composition, (ii) the initial main-sequence abundance 
of oxygen, and iii) the actual oxygen abundance in PNe since it may
differ significantly from the initial value due to dredge-up episodes
during the TP-AGB phase.
\end{itemize}

Overall, {\em our results explain in a very natural way 
the successful calibration of the PNLF using Cepheid distances} 
(Feldmeier et al.\ 1997; Jacoby 1996). In the list given by 
Jacoby (1996), 7 out of 8 galaxies in the sample are classified 
as spiral or irregular galaxies, only NGC5253 is of E/S0 type. 
However, it is well documented that this galaxy has undergone 
a recent starbust (see, e.g.\ Verma et al.\ 2003) and thus, 
as for all the other objects star formation has taken place 
during the last billion year, and all PNLFs reach $M^*$. 
In fact, the argument can be made even stronger and more general: 
in case there is a Cepheid distance to a galaxy, there must 
be a young population and thus the PNLF will reach the canonical 
cutoff brightness.

Furthermore, {\em our models do not support the 
idea of a population-invariant
PNLF cut-off}. Indeed, the suggestions for an age-invariant cut-off
poses a serious problem to the theory. In order to explain it,
{we may need to explore the properties of very low-$Z$ PNe,
or to invoke} a very different and difficult-to-imagine scenario
for the evolution of PNe and their progenitor stars than here 
considered.

Considering all these aspects, 
we find it worth questioning if the indications
for the cut-off invariance are really as solid as claimed. 
Until definitive evidence is not brought, and until there is not
a solid theoretical framework 
supporting such evidence, we suggest that the PNLF should 
still be considered as a potentially useful probe of
stellar populations in galaxies, rather than a standard candle.

%%%%%%%%%%%%%%%%%%%%%%%%%%%%%%%%%%%%%%%%%%%%%%%%%%%%%%%%

%%%%%%%%%%%%%%%%%%%%%%%%%%%%%%%%%%%%%%%%%%%%%%%%%%%%%%
\begin{acknowledgements}
This study is funded by a DAAD--Vigoni collaboration
project, and the Italian Ministry of University and Scientific Research
(MIUR). P.M. and L.G. thank the Max-Planck-Institut f\"ur
Astrophysik for the kind hospitality. 
\end{acknowledgements}

%%%%%%%%%%%%%%%%%%%%%%%%%%%%%%%%%%%%%%%%%%%%%%%%%%

\end{document}